\documentclass[journal,9pt]{IEEEtran}
\usepackage{cite}      
\usepackage{graphicx}  
\usepackage{balance}
\usepackage[]{subfigure}
\usepackage{url}
\usepackage{times}
\usepackage{amsmath}   
\usepackage{amsfonts,amsthm,amssymb}
\usepackage{mathtools}
\usepackage{nicefrac}
\usepackage[none]{hyphenat}
\usepackage{array}
\usepackage{fontenc}
\usepackage[linesnumbered,ruled,vlined]{algorithm2e}
\usepackage{color}
\usepackage[table]{xcolor}
\usepackage{etoolbox}
\usepackage{bm}
\usepackage{comment}
\usepackage{caption}
\usepackage{multicol}
\usepackage{multirow}
\usepackage{tikz}
\usepackage{textcomp}
\usepackage{hyperref}
\usepackage{lipsum}

\newcommand\copyrighttext{%
  \footnotesize \textcopyright This work has been submitted to the IEEE for possible publication. Copyright may be transferred without notice, after which this version may no longer be accessible.}
\newcommand\copyrightnotice{%
\begin{tikzpicture}[remember picture,overlay]
\node[anchor=south,yshift=10pt] at (current page.south) {\fbox{\parbox{\dimexpr\textwidth-\fboxsep-\fboxrule\relax}{\copyrighttext}}};
\end{tikzpicture}%
}

\DeclareMathOperator{\rb}{\frac{2r_b(t)}{c}}
\DeclareMathOperator{\rbn}{\frac{\tau_b(t)}{\delta t}}
\begin{document}
\setlength{\textfloatsep}{0pt}
\setlength{\intextsep}{10pt plus 2pt minus 2pt}

\title{Classification Of Automotive Targets Using Inverse Synthetic Aperture Radar Images}

\author{Neeraj~Pandey,~\IEEEmembership{Student Member, IEEE}, Shobha~Sundar~Ram,~\IEEEmembership{Senior Member, IEEE}
\thanks{N.P and S.S.R. are with the Indraprastha Institute of Information Technology Delhi, New Delhi 110020 India. E-mail: \{neerajp, shobha\}@iiitd.ac.in.}%
}

\maketitle
\copyrightnotice
\begin{abstract}
We present a framework for simulating realistic inverse synthetic aperture radar images of automotive targets at millimeter wave frequencies. The model incorporates radar scattering phenomenology of commonly found vehicles along with range-Doppler based clutter and receiver noise. These images provide insights into the physical dimensions of the target, the number of wheels and the trajectory undertaken by the target. The model is experimentally validated with measurement data gathered from an automotive radar. The images from the simulation database are subsequently classified using both traditional machine learning techniques as well as deep neural networks based on transfer learning. We show that the ISAR images offer a classification accuracy above 90\% and are robust to both noise and clutter.
\end{abstract}
\section{Introduction}
With the advent of advanced driver assistance systems (ADAS), automotive radars are becoming increasingly common on cars for improving road driving conditions. These radars are used for multiple applications such as automatic cruise control, pedestrian detection, cross-traffic alert, blind-spot detection, and parking assistance \cite{rasshofer2005automotive,schneider2005automotive}. The main advantage of automotive radar over camera for object detection and classification is that the radar can be operated in low light conditions and fog. Secondly, these sensors are typically cheaper than cameras and hence multiple of them can be mounted around the periphery of the vehicle, usually behind the bumpers.
Finally, automotive radars operate at millimeter wave frequencies with high bandwidths and spatially large antenna arrays. Hence, they offer excellent range, Doppler, and azimuth resolution \cite{hasch2012millimeter}. Usually, in these systems, the raw radar data cube is processed to provide a collection of point scatterers corresponding to both vehicles and clutter with range, azimuth, elevation, and Doppler information. Direct object detection and classification based on this type of data can be challenging since it is difficult to correctly cluster the point scatterers belonging to the same object \cite{fleming2012recent}. Instead, radar images / signatures directly processed from raw radar data provide more effective features for automatic target recognition. 

Classification of radar targets for a variety of application has been researched over the last few decades \cite{jouny1993classification,chiang2000model,bilik2006gmm}. Many different types of radar signatures have been studied. For example, radar micro-Doppler spectrograms, which are the joint time frequency representations of time-domain narrowband radar data, have proven to provide excellent features for classification. They have been used for distinguishing between different types of human activities \cite{kim2009human,kim2015human,vishwakarma2018dictionary,seyfiouglu2018deep}; armed and unarmed personnel \cite{fioranelli2015classification}; ground vehicles and pedestrians \cite{prophet2018pedestrian,khomchuk2016pedestrian}; and different types of airborne targets such as unmanned aerial vehicles and birds. \cite{ritchie2016multistatic,molchanov2013classification} for diverse applications like assisted living, elderly care and security, and surveillance. With broadband radar data from frequency modulated continuous waveforms, range-Doppler plots have been generated that have also served as excellent features for target recognition \cite{bilik2007radar}. Other works have used range-crossrange images generated through synthetic aperture radar (SAR) imaging for classification purposes \cite{chen2016target,wagner2016sar,lin2017deep}. However, SAR images are typically more suited for classifying static targets since dynamic targets may distort the radar images. The alternative is to use antenna array processing for obtaining fine cross-range resolution. However, this requires a large array with lots of antenna elements, and precise phase synchronization across the multiple channel data \cite{ram2015high}. A third method for obtaining fine cross-range resolution is to use inverse synthetic aperture radar (ISAR) processing of single-channel broadband data. 

When a dynamic target travels along a complex trajectory, the target undergoes a combination of translational and rotational motion. If the translational motion of a target can be correctly estimated and compensated, then the Doppler dimension can be mapped to cross-range to obtain ISAR images \cite{chen2014inverse}. The cross-range resolution is inversely proportional to the extent of the target aspect presented to the radar during the rotational motion and the coherent processing time interval. ISAR images, generated from range-Doppler plots of dynamic targets, have been extensively researched over the last two decades - especially for the detection and classification of airborne targets and ships \cite{martorella2009automatic,vespe2007automatic,park2011construction}.
In the automotive target scenario, vehicles undergo complex turns, which can result in a large radar aspect. Further, even while moving along a straight path, a slight offset of the target vehicular trajectory from the ego radar, can result in large radar aspects to get a fine cross-range resolution. Since automotive short range radars are characterized by large bandwidths (above 2GHz) and high carrier frequencies (77GHz) that result in fine range and Doppler (or cross-range) resolution, they are particularly suited for generating high resolution ISAR images of vehicles. In ADAS systems, multiple auxiliary sensors (gyrometers, accelerometers, other radars) are deployed on the ego vehicle. Therefore, the translational motion compensation of both the ego vehicle and target vehicle can usually be carried out without too much difficulty. More recently, ISAR images of ground based targets have been generated using turntable data \cite{danylov2010terahertz}, ground based platforms \cite{kulpa2013experimental,li2018wide}, and from airborne platforms \cite{essen2008high}. 
However, these studies have been restricted to very few targets. In our preliminary paper in \cite{pandey2020database}, we showcased how these images provide detailed insights into the dimensions of vehicles and their trajectories. However, the images were idealized and free of corruption from noise and ground clutter. 

Our contributions in this paper are as follows:
First, we provide a detailed simulation framework to generate realistic ISAR images of automotive targets while incorporating the effects of additive noise in the radar receiver and speckle noise due to ground clutter effects. The main objective is to provide a simulation framework for rapidly generating large volumes of radar data without the cost and man hours involved in collecting measurement data. These data can be used for training deep neural networks, which have recently emerged as the algorithm of choice for classifying radar images \cite{kim2015human,wagner2016sar,seyfiouglu2018deep}. Secondly, the simulation framework can be integrated with software test beds for rapid prototype development and validation. Finally, the simulation models can be useful for understanding radar propagation phenomenology in the environment and pin pointing cause and effect. 
We have considered five commonly found targets - a full-size car, a mid-size car, a four-wheel truck, a bicycle, and an auto-rickshaw (tuk-tuk). We generate ISAR radar images of these targets performing different types of turns (right, left, and U-turn) as well as following a straight trajectory. 

Second, in real world conditions, there can be considerable clutter arising from the rough road surface at millimeter wave frequencies, which is proportional to the radar coverage area \cite{skolnik1980introduction}. Hence, this range-based clutter cross-section increases with radar range. 
Further, Doppler based clutter can arise due to wind \cite{kulemin2003millimeter}. The combination of range and Doppler based clutter manifest as speckle noise and can significantly distort the ISAR images. In this work, we have incorporated detailed range-Doppler clutter as well as receiver noise in our radar models to simulate realistic radar images. With this paper, we release our database, consisting of over 30000 realistic ISAR images, to the radar community at \url{https://tinyurl.com/yyuwefrh}. 

Third, we demonstrate that these images show detailed information of the type of vehicle, its dimensions, the number of wheels and the trajectory followed by the vehicle. Further, we have validated these simulated images with measurement data gathered from Texas Instrument's AWR 1843 77GHz automotive radar. 

Finally, we demonstrate that the ISAR images offer distinctive features for the classification of automotive targets. In particular, we have considered both traditional machine learning algorithms such as support vector machine (SVM) \cite{suykens1999least} and random forest (RF) \cite{liaw2002classification} as well as Alexnet and Googlenet, which are two transfer learning based deep neural networks \cite{torrey2010transfer}. Our results show that the ISAR images are successfully classified by the machine learning algorithms successfully (with a precision and recall above 90\%). The deep neural networks outperform the traditional machine learning algorithms and robust to noise and clutter.  

The paper is organized as follows. In the Section \ref{sec:SimMethod} , we present the simulation methodology for modeling the scattered signal radar signals from the automotive targets, as well as the noise and clutter models. Then we describe the radar signal processing algorithms for generating the ISAR images. In Section \ref{sec:MeasSystem}, we present the experimental set up for collecting measurement data for generating ISAR images and present the measurement results. In Section \ref{sec:Results}, we present the classification results of the five automotive targets using four different machine learning based algorithms - SVM, RF, Alexnet and Googlenet. Finally, we conclude the paper with our final remarks in Section \ref{sec:Conclusion}.
\section{Simulation Methodology}
\label{sec:SimMethod}
While several prior works have described simulation models of pedestrians \cite{deep2020radar,ram2010simulation}, there are very few works that model automotive vehicles along complex trajectories \cite{duggal2019micro, duggal2020doppler}. These works have confined their scope to simulating high range resolution profiles and micro-Doppler spectrograms.
In this section, we discuss the simulation methodology for modeling the scattered radar signals from five common automotive targets - bicycle, auto-rickshaw, mid-size car, full-sized car, and truck. Then we describe how these models can be integrated into the radar waveform to obtain ISAR images. Finally, we present the method to incorporate noise and clutter in the images. 
\subsection{Automotive Animation Model}
\label{subsec:AnimationModel}
\begin{figure}[htbp]
    \centering
    \includegraphics[width=3.4in,height=2.8in]{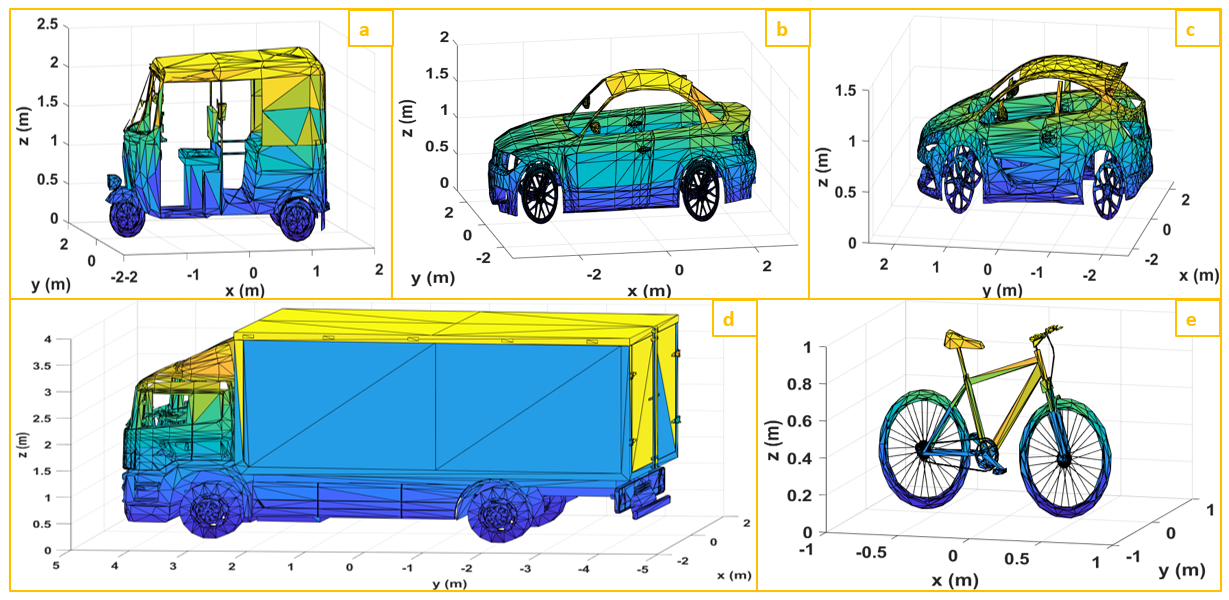}
    \caption{Three-dimensional automotive targets with triangular facets used for the simulation. (a) Auto-rickshaw (tuk-tuk), (b) full-sized car (c) mid-sized car, (d) truck (e) bicycle }
   \vspace{-2mm}
    \label{fig:Vehicles}
\end{figure}

\begin{figure}[htbp]
    \centering
    \subfigure[]{
    \includegraphics[width=1.6in,height=1.6in]{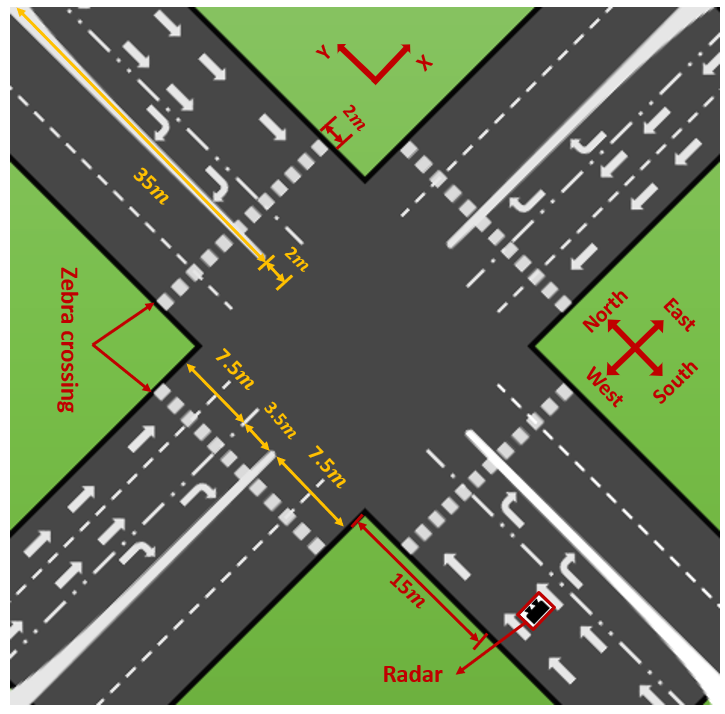}}
    \subfigure[]{
    \includegraphics[width=1.6in,height=1.6in]{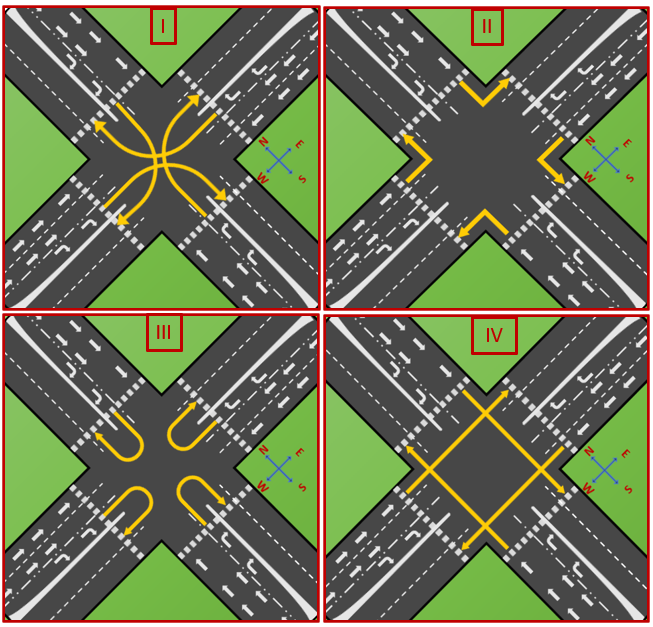}}
    \caption{(a) Road geometry of four-way traffic junction. (b) Trajectories undertaken by the automotive target in a four-way junction - (i) Right turn, (ii) Left turn, (iii) U-turn and (iv) Straight through.}
    \vspace{-2mm}
    \label{fig:trajectries}
\end{figure}

We imported freely available three-dimensional (3D) computer aided design (CAD) models\footnote{https://free3d.com/3d-models/} of the automotive targets into Blender software. Then, we rendered the metallic parts of the automotive into triangular facets. An accurate rendering of the target capturing the diversity of features on the chassis of the vehicle is realized by using a large number of facets. In our work, the bicycle and auto-rickshaw are rendered with 3919 and 6949 facets, respectively; the mid-size and full-size cars with 6905 and 19964 facets, respectively while the truck is rendered with 7206 facets as shown in Fig.\ref{fig:Vehicles}.
We have considered a four-way traffic junction, where lanes from the north (N), south (S), east (E), and west (W) meet as shown in Fig\ref{fig:trajectries}a. 
The targets are assumed to stand on the XY ground plane which is aligned with the N-S and E-W directions with the height along the Z-axis. The ego radar is assumed to be static and fixed at $(0,0,0.5)$m along the south road.
A total of 16 different trajectories are possible at this junction. They are the four right turns (S to E, E to N, N to W, W to S), four left turns (S to W, W to N, N to E, E to S), four U-turns (S to S, E to E, N to N, W to W) and four straight through (S to N, N to S, W to E, E to W) as shown in Fig\ref{fig:trajectries}b. 

\begin{figure}[htbp]
    \centering
    \subfigure[]{
    \includegraphics[width=1.6in,height=1.5in]{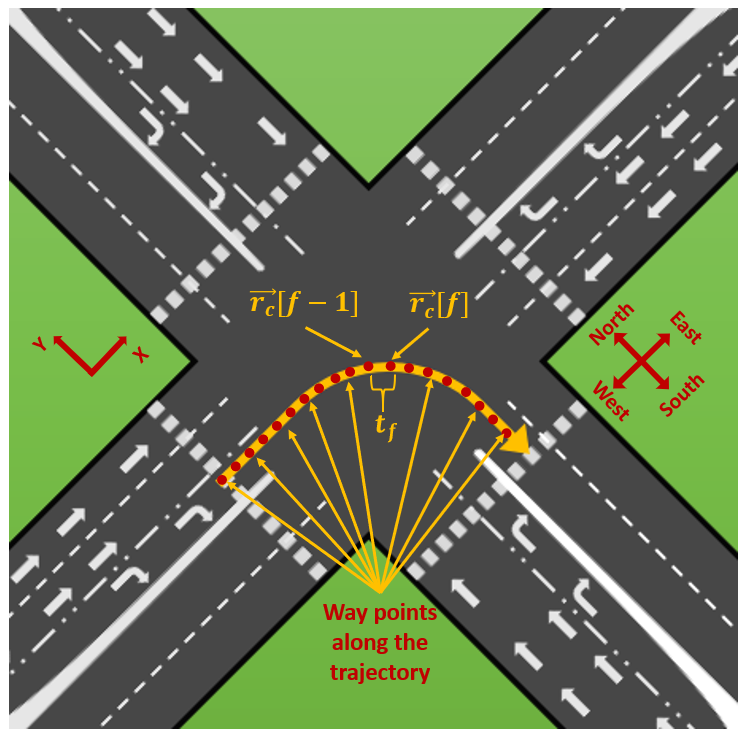}}
    \subfigure[]{
    \includegraphics[width=1.2in,height=1.2in]{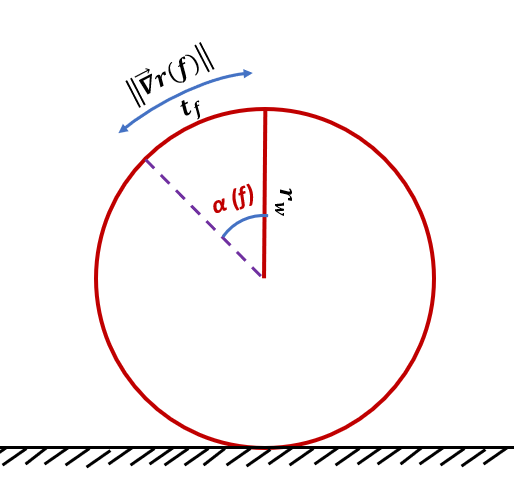}}
    \caption{Animation model of trajectory: (a) Way points along the trajectory that will be traversed by the center of the vehicle; (b) wheels rotational angle calculation.}
    \vspace{-2mm}
    \label{fig:Animation_model}
\end{figure}

Now, we describe the method for animating the vehicle along a desired trajectory at a specified speed. We first identify the center of the vehicle $\vec{r}_C$ and fix it at the starting position along a trajectory. Then, we identify way points along the distance of the trajectory such that the time taken for the vehicle's center to travel between any two way points is fixed ($t_f$) as shown in Fig.\ref{fig:Animation_model}a. 
The sampling time instants corresponding to these way points are indicated by $f=1,2,\cdots,F$. Therefore, the center of the vehicle undergoes translational displacement $\vec{\nabla}r[f]=\vec{r}_C[f]-\vec{r}_C[f-1]$ at each $f$. Now, the target is composed of $B$ triangular facets where the centroid of each facet is $\vec{r}_b, b = 1 :s B$. Based on the rendering of the vehicle, we obtain displacement vectors ($\vec{\nabla}r_b$) of these centroids from the center of the vehicle. These displacement vectors are fixed with time since the chassis of the vehicle behaves as a rigid body (excepting the wheels).
Now, the vehicle must undergo rotational motion along with translational motion. This rotational motion at each frame $f$ is described in terms of the yaw (rotation angle about the $Z$ axis, $\theta_b[f]$) which is computed from
\par\noindent\small
\begin{align}
    \theta[f] = \arctan\left(\frac{y_C[f]-y_C[f-1]}{x_C[f]-x_C[f-1]}\right.
\end{align}
The position of the facet centroid on the chassis of the vehicle is
\par\noindent\small
\begin{align}
    \vec{r}_b[f]  = \left(R_{\theta[f]}\vec{r}_b[f-1]\right) + \vec{\nabla}r[f],
\end{align}
where $R_{\theta}$ is the Euler rotation matrix for a yaw of $\theta$. 
In the case of wheels, a facet centroid on the wheel undergoes additional rotation due to the motion of the wheel. The angular displacement of the wheel is proportional to the distance travelled by the wheel and the radius of the wheel $r_w$ as shown in
\par\noindent\small
\begin{align}
    \alpha[f] = \frac{||\vec{\nabla}r[f]||}{r_w}.
\end{align} 
The axis of this wheel rotation is obtained by the cross product of the direction of translational displacement and the height axis. 
The total displacement of a point on the wheel is 
\par\noindent\small
\begin{align}
    \vec{r}_b[f]  = \left(R_{\alpha[f]}R_{\theta[f]}\vec{r}_b[f-1]\right) + \vec{\nabla}r[f],
\end{align}
where $R_{\alpha}$ is the Euler rotation matrix corresponding to a pitch angle of $\alpha$.
The entire algorithm describing the animation motion modeling is summarized in Algorithm.\ref{alg:AnimationModel}.
\begin{algorithm}
\SetAlgoLined
\KwData{Fixed displacement vectors corresponding to facet centroids on the chassis and wheels of the vehicle ($\vec{\nabla}r_b, b+1:B$) with respect to the center of the vehicle ($\vec{r}_C$).}
\KwData{Specify way point positions for center of vehicle along trajectory: $\vec{r}_C[f], f=1:F$. Time duration between two way points is fixed ($t_f$).}
\KwResult{Time-varying position coordinates of facet centroids on vehicle ($\vec{r}_b[f], b=1:B, f=1:F$)}

Initialization: Initialize positions of all the facet centroids
$\vec{r}_b[1] = \vec{r}_C[1]+\vec{\nabla}r_b$, $b = 1:B$\;
 \For{f=2:F}{
  Compute displacement between two consecutive way points $\vec{\nabla}r[f] = \vec{r}_C[f] - \vec{r}_C[f-1]$\;
  Compute vehicle yaw rotation:
  $\theta[f] = \frac{y_C[f]-y_C[f]}{x_C[f]-x_C[f-1]}$\;
  
  \uIf{Facet centroids on chassis}{
   Perform Euler rotation on facet centroids
   $\vec{r}_b[f] = \left(R_{\theta[f]}\vec{r}_b[f-1]\right)+\vec{\nabla}r[f], b=1:B$\;
   }
   \ElseIf{Facet centroids on wheels}{
   Compute rotation of wheel $\alpha[f] = \frac{||\vec{\nabla}r[f]||}{r_w}$ where $r_w$ is the radius of the wheel.\;
   Compute axis of wheel rotation which is perpendicular to the plane defined by height axis and vehicle translational motion direction.\;
   Perform Euler rotation on facet centroids on wheel based on wheel rotation axis
   $\vec{r}_b[f] = \left(R_{\alpha[f]}R_{\theta[f]}\vec{r}_b[f-1]\right)+\vec{\nabla}r[f], b=1:B$
  }
  \Else{}{}
 }
 \caption{Animation model of vehicle along desired trajectory }
 \label{alg:AnimationModel}
\end{algorithm}
\subsection{Electromagnetic Model of Radar Scattering}
\label{sec:ISARmodeling}
Automotive radars use linear frequency modulated (LFM) waveforms for detection and tracking of targets. We consider a radar transmitting an LFM waveform, 
\par\noindent\small
\begin{align}
\label{eq:TxSig}
    s^{tx}(\tau) = rect\left(\frac{\tau}{T_{PRI}} \right)e^{j2\pi f_c \tau}e^{j \pi K \tau^2},
\end{align}
with $f_c$ carrier frequency (and wavelength $\lambda_c$) and a chirp rate $K$.
In the above expression, $rect(\cdot)$ indicates that the transmitting signal is defined for a pulse repetition interval of $T_{PRI}$. We model the automotive target as a collection of moving scattering centers, $b = 1:B$, located at the centroids of each of the facets on the vehicle. The time-varying range for each scattering center is $r_b = R_b + v_bt$, where $R_b$ is the starting distance from the radar.
The radar signal scattered back from a single point target is Doppler shifted by $f_{D_b} = \frac{2v_b}{\lambda_c}$ due to the target's relative radial velocity ($v_b$) with respect to the radar. The received radar signal, after downconversion to the baseband, is written in terms of slow time $t$ and fast time, $\tau$ as 
\par\noindent\small
\begin{align}
\label{eq:RxSig}
\begin{split}
    s^{rx}_b(\tau,t) = a_b(t) rect\left( \frac{\tau - \tau_b}{T_{PRI}}\right)\\ e^{-j\frac{4\pi f_c}{c}R_{b}}e^{-j2\pi f_{D_b} t}e^{j \pi K \left(\tau - \tau_b\right)^2 } + \nu,
\end{split}
\end{align}
where $\tau_b(t)=\rb$ is the time delay to the target.
In the equation above, $\nu$ denotes the additive noise that will be discussed in greater detail in the following section. The strength of the received returns, denoted by $a_b$, is obtained through the radar range equation by incorporating the transmitted power ($P^{tx}$), the gains of the transmitting ($G^{tx}$) and receiving radar antennas ($G^{rx}$), the radar cross-section of each scatterering center ($\sigma_b$) and the distance of the point scatterer from the radar, as shown in
\par\noindent\small
\begin{align}
\label{eq:RadarRangeEq}
    a_b^2 = \frac{P^{tx} G^{tx}G^{rx} \sigma_b \lambda_c^2}{(4\pi)^3 r_b^4}.  
\end{align}
In \eqref{eq:RxSig} and \eqref{eq:RadarRangeEq}, we have assumed stationary channel conditions and direct path target returns without any type of multipath. The RCS of a flat metallic triangular plate is a function of the radar aspect angle ($\theta_b$), the plate area ($A_b$) and long dimension ($d_b$), \cite{ruck1970radar}, as shown in 
\par\noindent\small
\begin{align}
\label{eq:rcs_triangle}
\sigma_b = \eta\frac{4\pi A_b^2 \cos^2 \theta_b}{\lambda_c^2}\frac{\sin^4 \left(\frac{2\pi}{\lambda_c}d_b \sin \theta_b \right)}{\left(\frac{2\pi}{\lambda_c}d_b \sin \theta_b\right)^4}.
\end{align}
The aspect angle is computed from the dot product of the incident vector from the radar to the plate and the normal vector of the plate. 
Not all scattering centers may be visible to the radar. Since $\theta_b$ changes along the target trajectory, $\sigma_b$ fluctuates. Hence, we incorporate a Bernoulli's random variable, $\eta$, with a 50\% probability of visibility to the radar.
The radar data is sampled at a frequency of $F_s = 1/\delta t$ and the fast time samples are numbered from $1:N$. Similarly, if we consider a $p^{th}$ coherent processing interval (CPI) consisting of $M$ PRIs, then the discrete representation of \eqref{eq:RxSig} is
\par\noindent\small
\begin{align}
\begin{split}
       S^{rx}_b[n,m] = a_b[m] rect\left[ \frac{n - n_b}{N}\right] e^{-j\frac{4\pi f_c}{c}R_{b}} \\
       e^{-j2\pi m f_{D_b} T_{PRI}}e^{j \pi K \delta t^2 \left(n - n_b \right)^2 } + \nu,
\end{split}
\end{align}
where $n_b$ is the integer rounded from $\rbn$. 

We process the received signal using \emph{stretch processing}, a variation of matched filtering which is especially suited for low sampling frequencies \cite{richards2005fundamentals}. The maximum unambiguous range of the radar, $R_{max}$, is equal to $\frac{c T_{PRI}}{2}$. For every CPI, we consider a radar range span of interest, $R_0 - \frac{R_{span}}{2}:R_0+\frac{R_{span}}{2}$ within $R_{max}$ where $R_0$ is called the central reference position (CRP). The time delay to the $b^{th}$ point scatterer can be expressed as $\tau_b = \tau_0 + \delta \tau_b(t)$ where $\tau_0 = \frac{2R_0}{c}$ corresponds to the time delay to the CRP. Since, the target motion is known, the CRP is chosen to correspond to the mean range to $\vec{r}_C$ in every CPI. 
In stretch processing, the received signal is multiplied with $e^{-j \pi K \delta t^2(n-n_0)^2}$, where $n_0$ is the integer rounded from $\frac{\tau_0}{\delta t}$ over every PRI. Thus, we obtain 
\par\noindent\small
\begin{align}
\begin{split}
\label{eq:StretchProcessorOutput}
     S^{rx}_b[n,m] = a_b[m] e^{-j\frac{4\pi f_c}{c}R_{b}}e^{-\pi K \delta_t^2 (n_0-n_b)^2} e^{-j2\pi m f_{D_b} T_{PRI}}\\ e^{-j4\pi K \delta_t^2 n (n_0-n_b) } .
\end{split}
\end{align}
We carefully compensate for the translational motion of the vehicle, and only consider the rotational motion of the point scatterer within a CPI. Then, the first two exponential terms in \eqref{eq:StretchProcessorOutput} are constant phase terms and are absorbed into the amplitude during further processing.
The last two terms show the variation of the two-dimensional (2D) signal over slow and fast times, as shown in
\par\noindent\small
\begin{align}
\label{eq:ExtendedTarget}
    S^{rx}_b[m,n] = a_b(\cdot)e^{-j 2\pi f_{D_b} mT_{PRI}} e^{-j 4\pi K \delta t^2 n (n_0-n_b) }.
\end{align}
The fast time sampling frequency $(F_s = 1/\delta t)$ is obtained from twice the stretch bandwidth which is $\frac{2R_{span}K}{c}$ where $R_{span}$  is much lower than $R_{max}$. Therefore, stretch processing results in lower sampling frequency requirements than ordinary matched filtering where the sampling frequency is atleast twice the bandwidth of the transmitted signal ($K T_{PRI} = \frac{2R_{max} K}{c}$).

When the target is an extended target with multiple point scatterers ($B$), then the received signal is obtained by the sum of the returns from each scatterer. 
\par\noindent\small
\begin{align}
  S^{rx}[m,n] = \sum_{b=1}^BS^{rx}_b[m,n].
\end{align}
Here, we have ignored the multiple scattering between the different parts of the target. 
The output is processed using 2D Fourier transform to obtain range-Doppler ambiguity plots, 
\par\noindent\small
\begin{align}
\label{eq:ISARimage_clean}
    \chi[f_D,r] = \mathcal{DFT}_{2D}\{S^{rx}[m,n]\},
\end{align}
where the range dimension $r$ spans $N$ steps from $R_0-\frac{R_{span}}{2}$ to $R_0 + \frac{R_{span}}{2}$; and $f_D$ spans $M$ steps from $-\frac{1}{2T_{PRI}}$ to $\frac{1}{2T_{PRI}}$. The 2D plot can also be interpreted as a range-cross plot $(\chi[r,cr])$ provided an accurate estimate of the angular velocity ($\omega$) of the target is available, since translational motion has been compensated. We estimate $\omega$ for every $p^{th}$ CPI by the change in yaw ($\Theta$) of the vehicle as shown in
\par\noindent\small
\begin{align}
    \omega = \frac{\Theta[p]-\Theta[p-1]}{T_{CPI}}.
\end{align}
Then the Doppler axis is converted to the cross-range axis by 
\begin{align}
    cr[m] = f_D[m] \times \frac{\lambda_c}{2 \omega}, \text{for } m = 0: M-1.
\end{align}
Depending on $\omega$, the cross-range spans across images may vary even when the pixel dimensions of the plot remain unchanged. 
\begin{table}[htp]
    \caption{Automotive radar parameters for generating ISAR images}
    \centering
    \begin{tabular}{p{3.5cm}|p{2cm}|p{2cm}}
    \hline \hline
        Parameters & Simulation & Measurement  \\
    \hline \hline
         Carrier frequency ($2\pi f_c$) & 77GHz & 77GHz\\ 
        stretch Bandwidth ($\frac{1}{T_{SBW}}$) & 8MHz & 2GHz \\
        Sampling Frequency ($F_s$) & 5MHz & 5MHz \\
        Chirp rate ($K$) & $60 \times 10^{12}$ Hz$^2$ & $7.5\times10^{12}$ Hz$^2$\\
        Chirp duration ($T_{PRI}$) & 83.33$\mu$s & 400$\mu$s\\
        Coherent processing interval ($T_{CPI}$) & 0.1s & 0.1s\\
        Transmitted power ($P_t$) & 25dBm & 14dBm\\
    \hline \hline
    \end{tabular}
    \vspace{-2mm}
    \label{tab:RadarParameters}
\end{table}
\subsection{Noise and Range-Doppler Clutter Models}
\label{subsec:RangeDopplerclutter}
In this section, we discuss how we incorporated ground based clutter along the range and wind based clutter along the Doppler dimensions along with additive noise in the time-domain data. 

\emph{Ground clutter:} For a rough surface, the clutter cross-section is proportional to the surface clutter coefficient, $\sigma^{0}$, and the radar coverage area. A stable component - due to static road conditions such as road material - and a fluctuating component due to wind contribute to $\sigma^{0}$ \cite{kulemin2003millimeter,ulaby2019handbook}:
\par\noindent\small
\begin{align}
\label{es:normalised RCS}
      \sigma^0=\sigma^0_{stable} + \sigma^0_{fluctuating}.
\end{align}
We model $\sigma^{0}$ as an exponential random variable with a mean of -15dB which corresponds to asphalt at millimeter wave frequencies \cite{king1970terrain}.
The radar coverage area is proportional to the antenna beamwidth ($\theta_{BW}$), grazing angle ($\psi$), radar range resolution ($\delta r=\frac{c}{2KT_{PRI}}$) and range. Therefore, $\sigma_c$ is 
\par\noindent\small
\begin{align}
    \sigma_{c} = \sigma^0 r \theta_{BW} \delta r \sec \psi.
\end{align}
This results in a ground clutter that is a function of range as shown in
\par\noindent\small
\begin{align}
\label{es:surface_clutter echo power}
      C_0[r]= \frac{P^{tx} G^{tx}G^{rx}  \sigma^0 \theta_{BW} \delta r \sec\psi}{(4 \pi)^2 r^3}.
\end{align}
We have maintained the radar at a height of 0.5m above the ground from which the grazing angle can be computed for every range $r$.

\emph{Doppler clutter:} 
Based on \cite{kulemin2003millimeter}, the power spectrum of the Doppler clutter can be modeled as a low pass filter response. When combined with the range related clutter, we obtain
\par\noindent\small
\begin{align}
\label{eq:DoppPowerSpectrum}
      C[f_D,r]= C_0[r]\left [1+\left(\frac{f_D}{\Delta f_D} \right)^{s}. \right ]^{-1}
\end{align}
where $s$ is a function of the average wind velocity ($U$) as shown in
\par\noindent\small
\begin{align}
\label{eq:nEqnDopp}
      s=\frac{2(U+2)}{(U+1)}\left(\frac{100}{2\pi f_c} \right )^{0.2}.
\end{align}
In \eqref{eq:DoppPowerSpectrum}, $\Delta f_D$ is the $-3dB$ width of the spectrum and is given by
\par\noindent\small
\begin{align}
\label{eq:3dBwidth}
      \Delta f_D= 1.23 \left(\frac{3.2}{\lambda_c} \right) U^{1.3}.
\end{align}
Based on local meteorological reports \cite{pandey2012study}, $U$ can vary from 0m/s to 10m/s. We consider four possible wind speeds ($2.5, 5, 7.5,10 $m/s) in our simulations. 
Finally, we convert the power values obtained from \eqref{eq:DoppPowerSpectrum} to voltage values for each range-Doppler pixel ($c[f_D,r]$). We multiply the voltage with a phase modeled as a complex circularly symmetric random variable ($\phi[f_D,r]$). This complex clutter signal ($c[f_D,r] exp(j\phi[f_D,r])$) is then added to each pixel of the range-Doppler ISAR images $\chi[f_D,r]$ in \eqref{eq:ISARimage_clean}.

\emph{Noise:} While clutter was modeled as a speckle noise in the radar images based on the above description, we modeled noise as an additive white Gaussian $\mathcal{N}(0,N_p)$ in the time-domain radar returns in \eqref{eq:RxSig}. Based on our radar range equation in \eqref{eq:RadarRangeEq}, the minimum received signal at the radar is $-80$ dBm. Based on the ratio between the minimum received signal at the radar and the mean noise floor, we considered four different signal to noise ratio (SNR) scenarios from $-5dB$ to $+10dB$ in our simulations. 

\subsection{Database of Simulated Data}
We present examples of ISAR images of each of the targets below. For all five automotive targets (bicycle, auto-rickshaw, mid-size car, full-size car, and truck), we present two sets of results. The figures on the left show the ISAR images corrupted by additive noise in the receiver data due to receiver electronics. We present the set of images corresponding to an SNR of $+10dB$. The figures on the right show the ISAR images corrupted by range-Doppler clutter that give rise to speckle noise. The clutter strength, in this scenario, is a function of the surface reflection coefficient of the road and the wind speed. We present figures corresponding to a wind speed of 2.5m/s. In all of the figures, each row corresponds to images simulated for a distinct trajectory. The top row is obtained when the vehicle moves along an almost straight trajectory from N to S; the second row shows the trajectory of a target taking a left turn from E to S; the third row shows the trajectory of a target taking a right turn from S to E; while the fourth row shows the trajectory of a target doing a U-turn from W to W. The range span (Y-axis) in all the figures is 20m and is centered along the CRP corresponding to the specific CPI. The cross-range span (X-axis) in all of the figures may vary from 10m to 20m and is centered at 0m. The cross-range axis correlates to the Doppler axis of $\chi[f_D,r]$. The noisy images on the left are of a dynamic range of $50dB$ from $-40dBm$ to $-90dBm$. While, the images on the right are of a dynamic range of 80dB from $-40dBm$ to $-120dBm$.

\begin{figure*}[htbp]
    \centering
    \subfigure[]{
    \includegraphics[ width=2.8in,height=2.8in ]{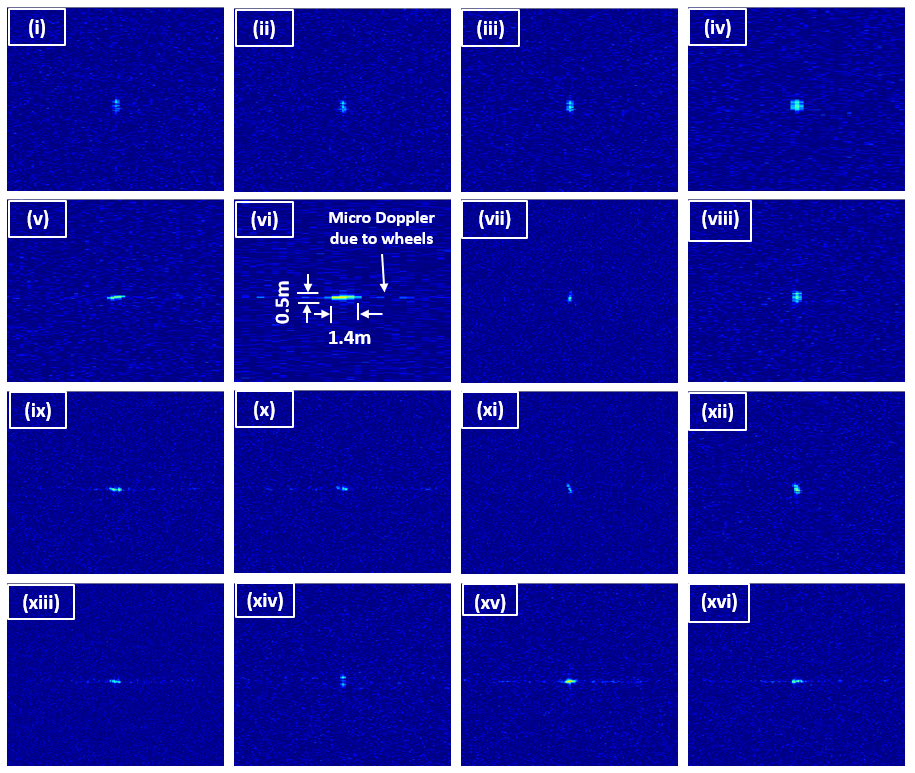}}
    \subfigure[]{    \includegraphics[ width=2.8in,height=2.8in ]{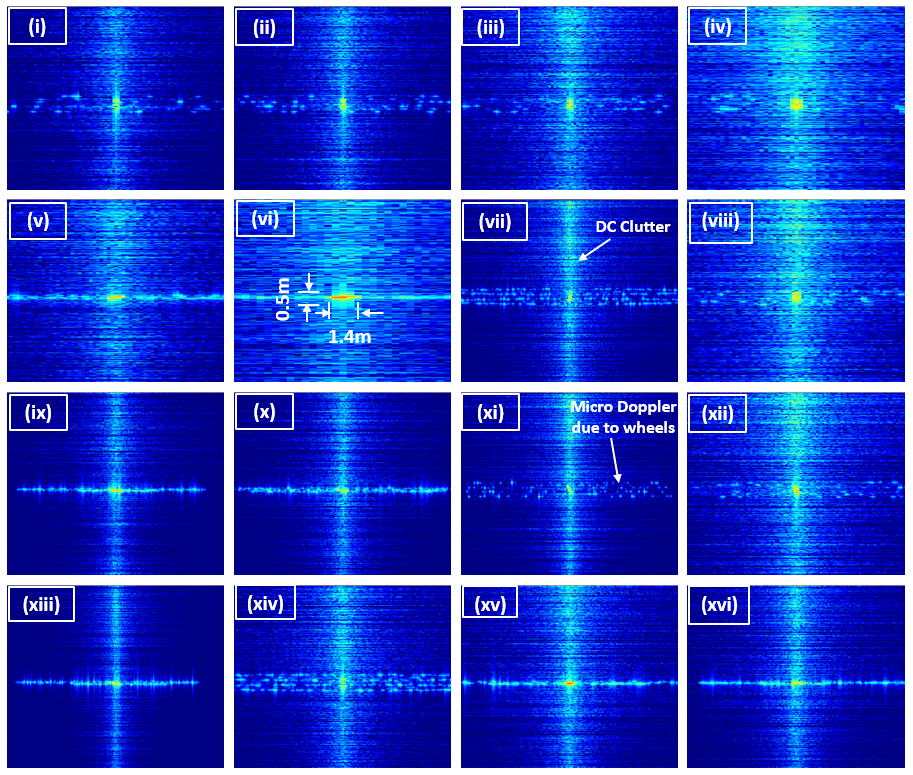}}
    \caption{ISAR images of a \emph{bicycle} of (a) +10dB SNR, (b) with range-Doppler clutter with wind velocity 2.5 m/s at CPI frames corresponding to 1.5, 2.5, 3.5, 4.5s along following trajectories: (i-iv) straight path from north to south, (v-viii) left turn from east to south (ix-xii) right turn from west to south and (xiii-xvi) U-turn from west to west. The range span is 20m while the cross-range span varies from 10m to 20m. The dynamic range for SNR is 50dB (-40dBm to -90dBm) and for clutter 80dB (-40dBm to -120dBm).}
    \vspace{-2mm}
    \label{fig:Bicycle_results}
\end{figure*}

We, first, present the ISAR images of a \emph{bicycle} in Fig.\ref{fig:Bicycle_results}. As mentioned earlier, the images in Fig.\ref{fig:Bicycle_results}a correspond to the ISAR images corrupted by noise while the figures on the right correspond to the ISAR images corrupted by range-Doppler based clutter.
The bicycle is a spatially narrow target and hence appears as a cluster of very closely spaced scattering centers almost like a single point scatterer. The dimensions of the target can be estimated from some of the images (for example, sub-figure vi). The noisy images on the left show that at some range positions, the target becomes difficult to discern due to low returns from the bicycle, especially when it is far from the radar. This is because of the low RCS of the bicycle. The cluttered images on the right show strong clutter at DC (corresponding to 0m along the cross-range). The width of the Doppler spectrum and the strength of the clutter returns change depending on the wind speed. The bicycle can still be discerned in some of the images along with the micro-Doppler tracks due to its wheel (sub-figures xi and xiv in (b)). 

\begin{figure*}[htbp]
    \centering
    \subfigure[]{
    \includegraphics[ width=2.8in,height=2.8in ]{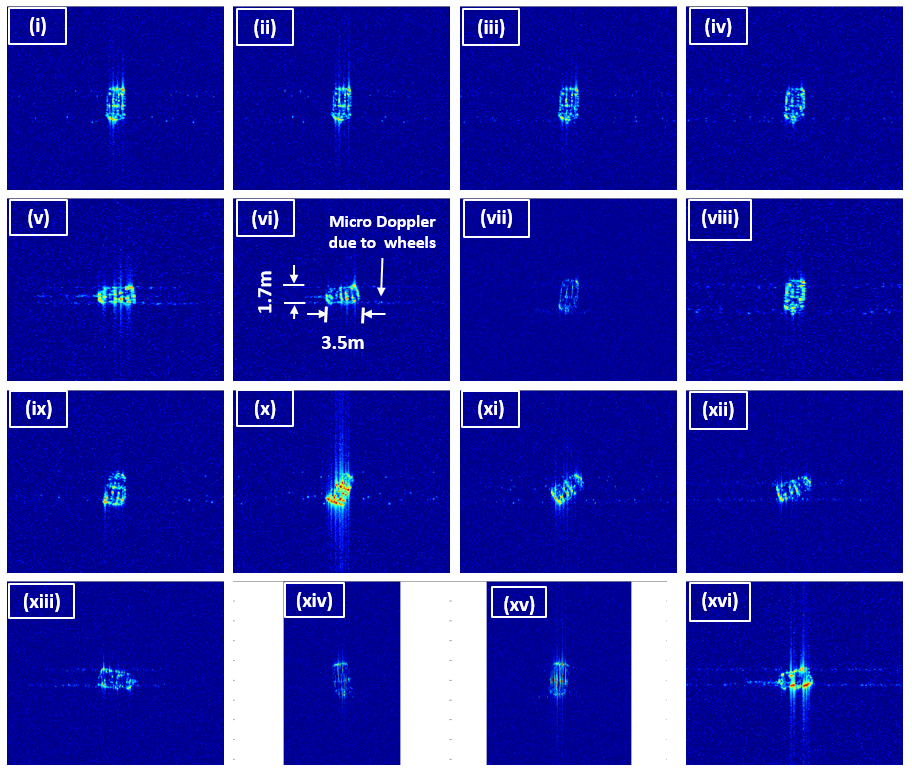}}
    \subfigure[]{    \includegraphics[ width=2.8in,height=2.8in ]{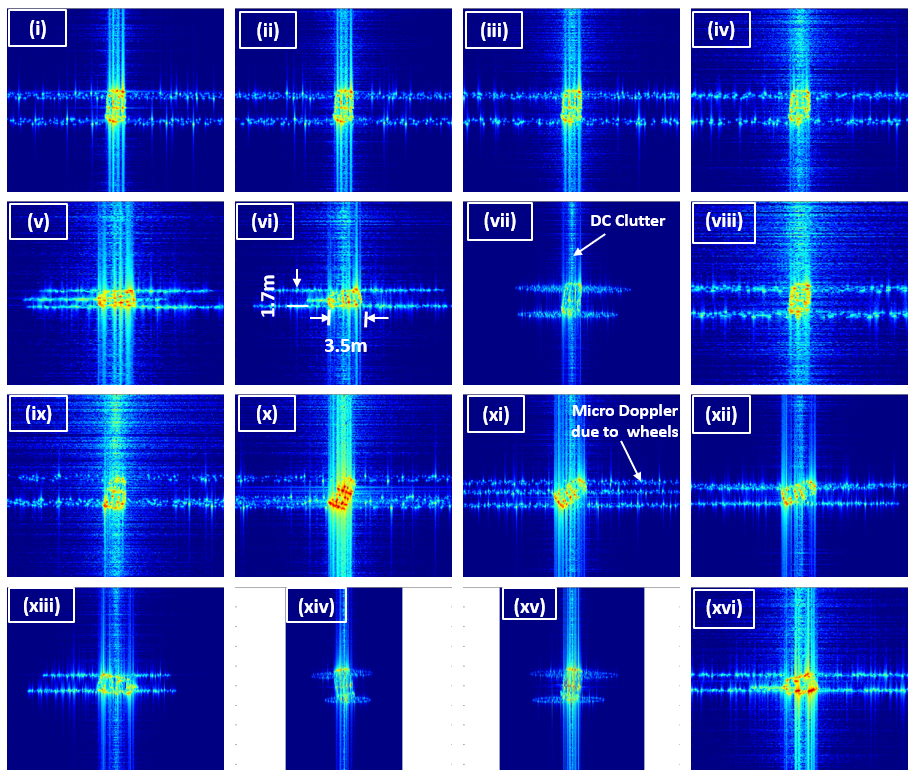}}
    \caption{ISAR images of an \emph{auto-rickshaw} of (a) +10dB SNR (b) with range-Doppler clutter with wind velocity 2.5 m/s at CPI frames corresponding to 1, 2, 3, 4s along following trajectories: (i-iv) straight path from north to south, (v-viii) left turn from east to south (ix-xii) right turn from south to east and (xiii-xvi) U-turn from west to west. The range span is 20m while the cross-range span varies from 10m to 20m. The dynamic range for SNR is 50dB (-40dBm to -90dBm) and for clutter 80dB (-40dBm to -120dBm). }
    \vspace{-2mm}
    \label{fig:Autorickshaw_results}
\end{figure*}

Next, we present the ISAR images of an \emph{auto-rickshaw} in Fig.\ref{fig:Autorickshaw_results}. 
The images show that the auto-rickshaw is a spatially larger target than the bicycle. The shape of the vehicle is triangular in the top-view. In fact, in some of the top-view images (sub-figure vi), we can clearly see the dimensions of the vehicle. We also see considerable distortions along the Doppler (cross-range) dimension due to micro-Doppler from the rotation of the wheels. Interestingly, in some images, we can see three distinct micro-Doppler tracks from the three wheels (sub-figure xi in (b)). On the top row, we observe that the longer dimension of the target's top view is oriented along the range dimension when the car moves from N to S. In the second row, the target is first oriented laterally and then turns length wise. This is because the target did a left turn from E to S. Similarly, in the third row, the target was first oriented along the long direction and then along the lateral direction as the target moved from S to E. Finally, in the last row, the target is always along the lateral direction since it does a U-turn from W to W. Therefore, the ISAR images offer some indication of the type of trajectory undertaken by a target.

\begin{figure*}[htbp]
    \centering
    \subfigure[]{
    \includegraphics[ width=2.8in,height=2.8in ]{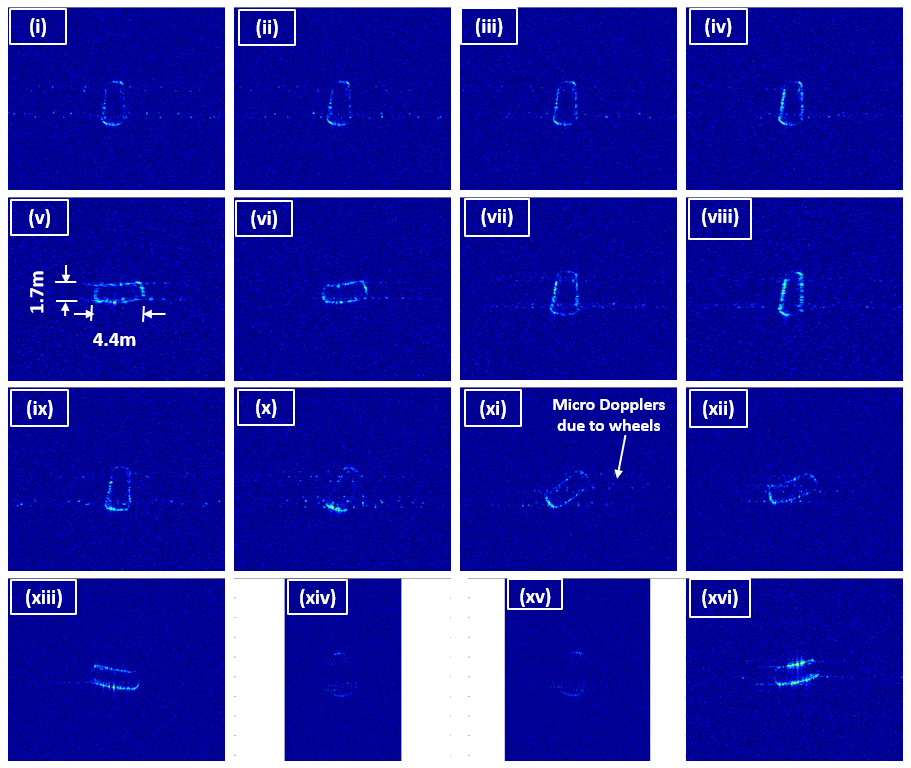}}
    \subfigure[]{    \includegraphics[ width=2.8in,height=2.8in ]{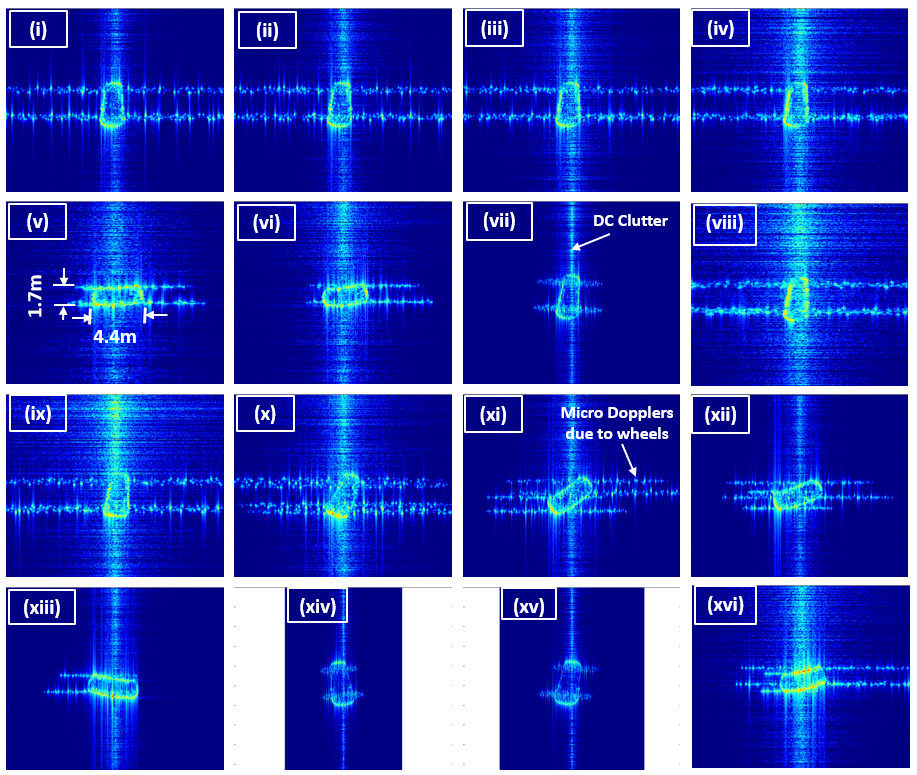}}
    \caption{ISAR images of \emph{mid-size car} of (a) +10dB SNR, (b) with range-Doppler clutter with wind velocity 2.5 m/s at CPI frames corresponding to 1, 2, 3, 4s along following trajectories: (i-iv) straight path from north to south,(v-viii) left turn from east to south (ix-xii) right turn from south to east and (xiii-xvi) U-turn from west to west. The range span is 20m while the cross-range span varies from 10m to 20m. The dynamic range for SNR is 50dB (-40dBm to -90dBm) and for clutter 80dB (-40dBm to -120dBm). }
    \vspace{-2mm}
    \label{fig:Midsize_car_results}
\end{figure*}

Figure.\ref{fig:Midsize_car_results} presents the results of the mid-size car.
Since this is a larger target than the auto-rickshaw, the returns are stronger. We are able to see the top view of the target with all four sides. Again, we observe some micro-Doppler based distortions along the cross-range due to the micro-Doppler returns from the four wheels. Four distinct micro-Doppler tracks are observed in the sub-figure x and xi in Fig.\ref{fig:Midsize_car_results}b. Again, we observe the longer dimension of the car oriented along the range dimension when it is moving from either N to S or S to N. But the longer dimension of the car is oriented along the lateral dimension when the car is moving from E to W or vice versa.

\begin{figure*}[htbp]
    \centering
    \subfigure[]{
    \includegraphics[ width=2.8in,height=2.8in ]{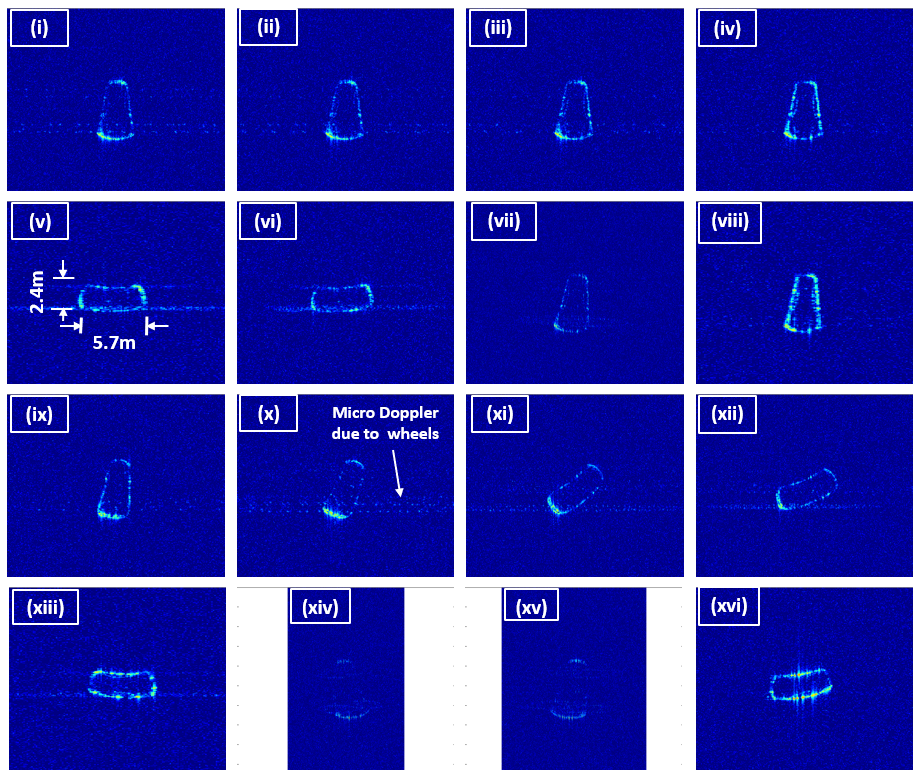}}
    \subfigure[]{
    \includegraphics[ width=2.8in,height=2.8in ]{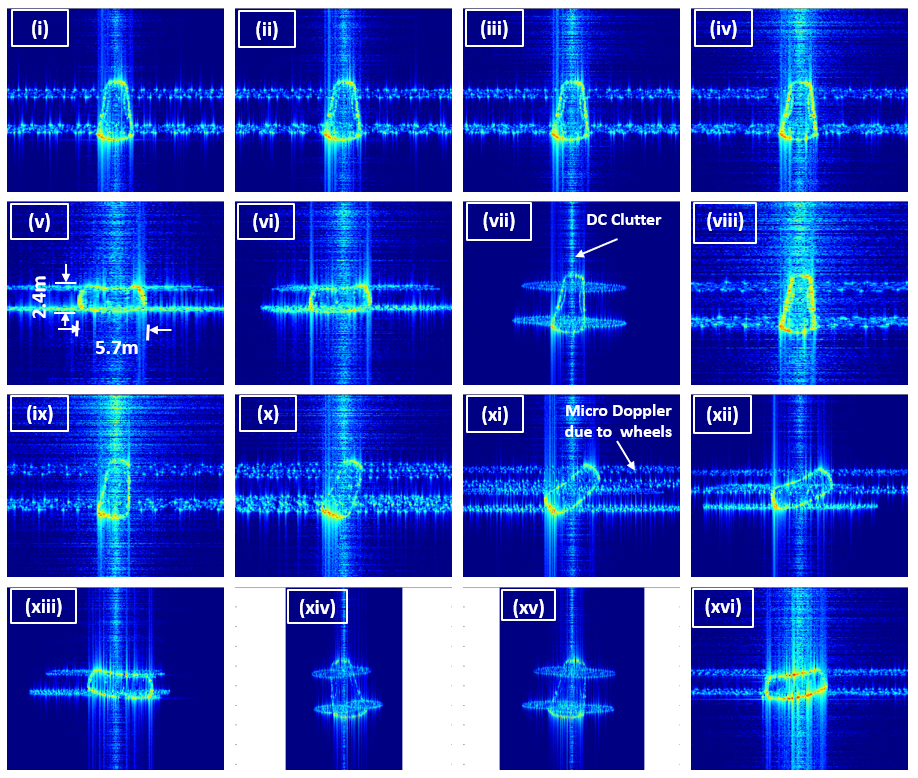}}
    \caption{ISAR images of \emph{full-size car} of (a)+10dB SNR, (b) with range-Doppler clutter with wind velocity 2.5 m/s, at CPI frames corresponding to 1, 2, 3, 4s along following trajectories: (i-iv) straight path from north to south, (v-viii) left turn from east to south (ix-xii) right turn from south to east and (xiii-xvi) U-turn from west to west. The range span is 20m while the cross-range span varies from 10m to 20m.The dynamic range for SNR is 50dB (-40dBm to -90dBm) and for clutter 80dB (-40dBm to -120dBm).}
    \vspace{-2mm}
    \label{fig:Fullsize_car_results}
\end{figure*}

The results in Fig.\ref{fig:Fullsize_car_results}, corresponding to the \emph{full size car} look similar to those from the mid-size car in Fig.\ref{fig:Midsize_car_results}, except for the larger dimensions of the car in the top-view. The dimensions of the full size car are 5.7m $\times$ 2.4m whereas those of the mid-size car were 4.4m $\times$ 1.7m.
The larger target also has stronger returns and is thus easily discerned in the images. Again, we are able to observe four distinct micro-Doppler tracks from the four wheels in some of the images (sub-figures ix-xii in Fig.\ref{fig:Fullsize_car_results}b). Also, we are able to see the changes in the orientation of the images as the car undergoes turns along its trajectory.

\begin{figure*}[htbp]
    \centering
    \subfigure[]{
    \includegraphics[ width=2.8in,height=2.8in ]{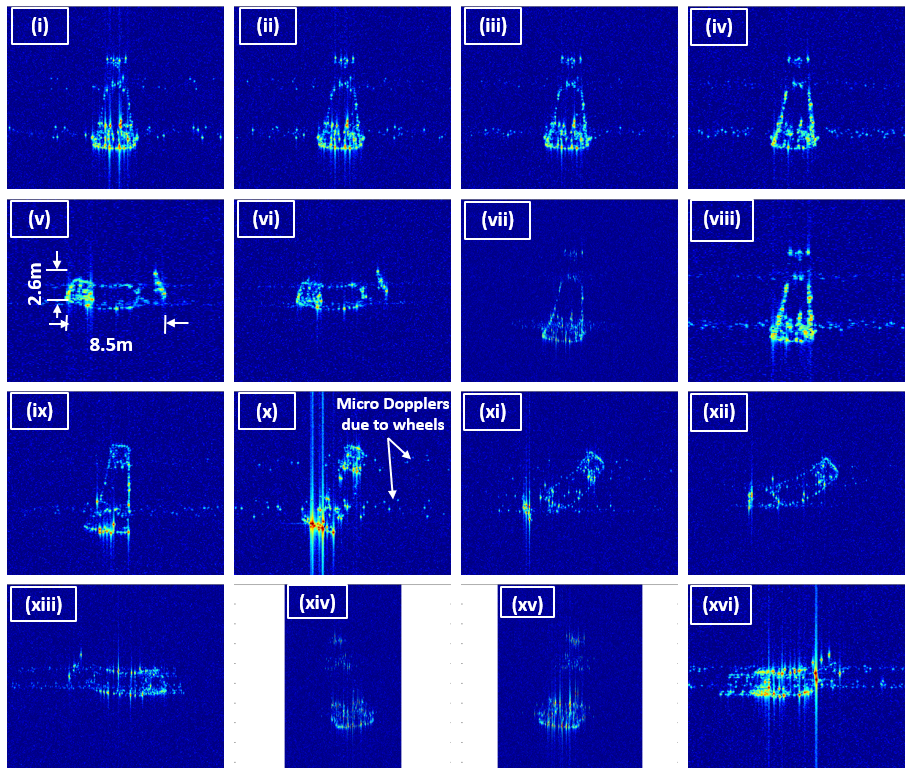}}
    \subfigure[]{    \includegraphics[ width=2.8in,height=2.8in ]{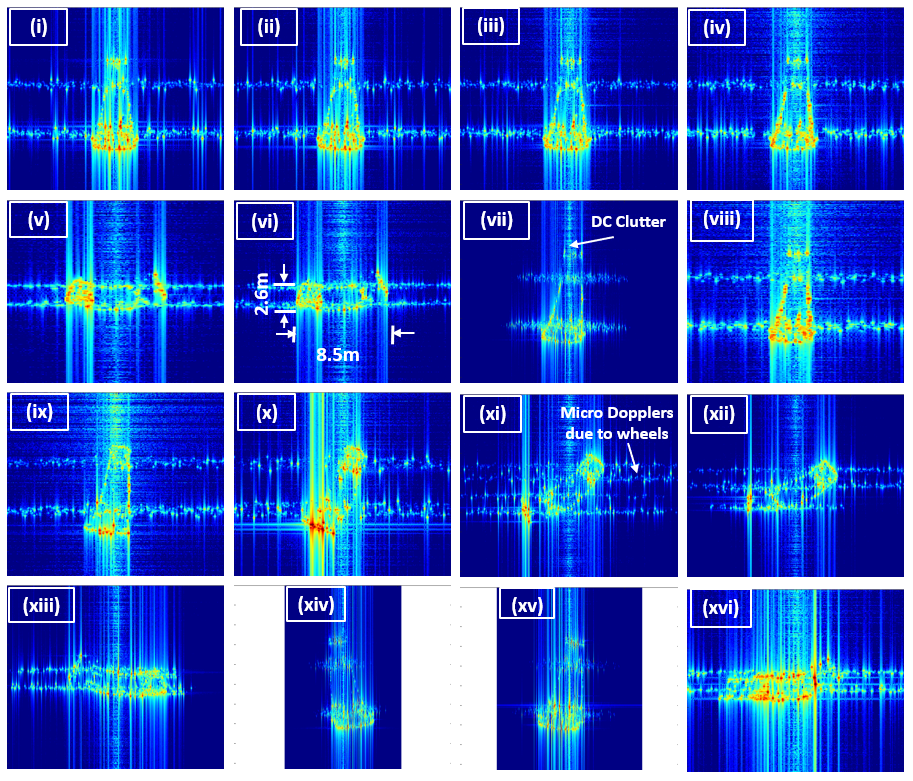}}
    \caption{ISAR images of \emph{truck} of (a)+10dB SNR, (b) with range-Doppler clutter with wind velocity 2.5 m/s, at CPI frames corresponding to 1, 2, 3, 4s along following trajectories: (i-iv) straight path from north to south, (v-viii) left turn from east to south (ix-xii) right turn from south to east and (xiii-xvi) U-turn from west to west. The range span is 20m while the cross-range span varies from 10m to 20m. The dynamic range for SNR is 50dB (-40dBm to -90dBm) and for clutter 80dB (-40dBm to -120dBm).}
    \vspace{-2mm}
    \label{fig:Truck_results}
\end{figure*}

The largest automotive target that we have considered is the \emph{four wheel truck}, for which, the results are presented in Fig.\ref{fig:Truck_results}.
Due to its large size, the top-view obtained from the ISAR images clearly present the dimensions of the target which are 8.5m $\times$ 2.6m. We are also able to observe the changes in the target orientation along the four distinct trajectories. The micro-Doppler distortions are considerably greater in this case due to the large wheel radii and four micro-Doppler tracks in sub-figures xi in Fig.\ref{fig:Truck_results} are well resolved in this case. 

We list the complete set of simulated ISAR images in Table.\ref{tab:ISAR_database}. 
\begin{table}[htbp]
    \centering
\caption{Simulated ISAR image database}
\begin{tabular}{ p{2.4cm}|p{1cm}|p{1.2cm}|p{1.1cm}|p{1cm}} 
\hline \hline
Type of images & Types of Targets ($\#$) & Trajectories ($\#$) & Images per trajectories & Total images ($\#$)\\
\hline \hline
Ideal Images & 5 & 16 & 45-49 & 3750\\
\hline
Noisy images of SNR (+10,+5,0,-5 dB) & 5 & 16 & 45-49 & 14976\\
\hline
Cluttered Images with wind velocities (2.5,5,7.5,10 m/s) & 5 & 16 & 45-49 & 14976\\
\hline \hline
\end{tabular}
\vspace{-2mm}
\label{tab:ISAR_database}
\end{table}
To summarize, we have considered five automotive targets - full-size car, mid-size car, truck, auto-rickshaw, and bicycle. Each target undergoes 16 trajectories, and each trajectory is of 5 seconds duration. Since each CPI is 0.1 seconds, we obtain between 45 and 49 images from each trajectory. Therefore, we obtain 3750 clean images that are free of noise and clutter. Then we corrupt these images with additive white Gaussian noise in the time-domain to obtain noisy images with SNR ranging from $-5$ to $+10$dB resulting in 14976 noisy images. Similarly, we introduce range-Doppler clutter with four different wind speeds ($U$) ranging from 2.5 m/s to 10 m/s to obtain 14976 cluttered images. With this paper, we publicly release this data set to the research community at \url{https://tinyurl.com/yyuwefrh}.
\section{Measurement Data}
\label{sec:MeasSystem}
\begin{figure*}[htbp]
    \centering
    \subfigure[]{
    \includegraphics[width=1.5in, height = 1.5in]{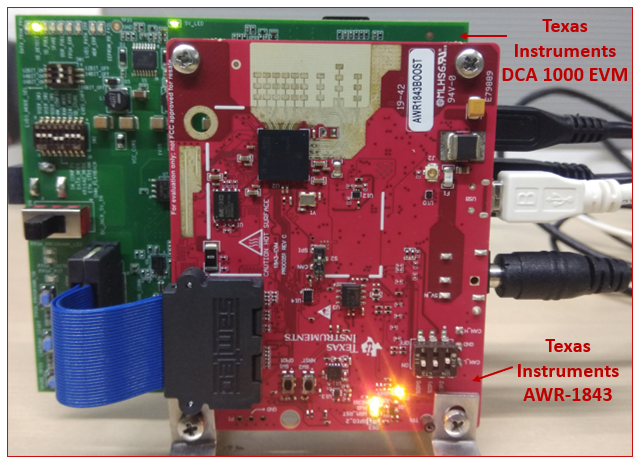}}
    \subfigure[]{
    \includegraphics[width=1.5in, height = 1.5in]{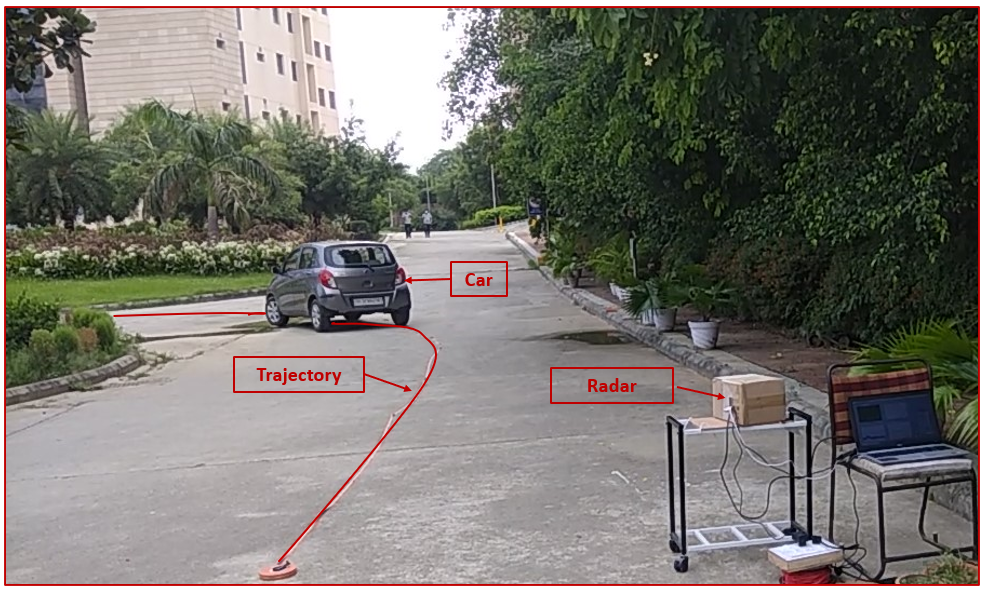}}
    \subfigure[]{
    \includegraphics[width = 1.5in, height = 1.5in]{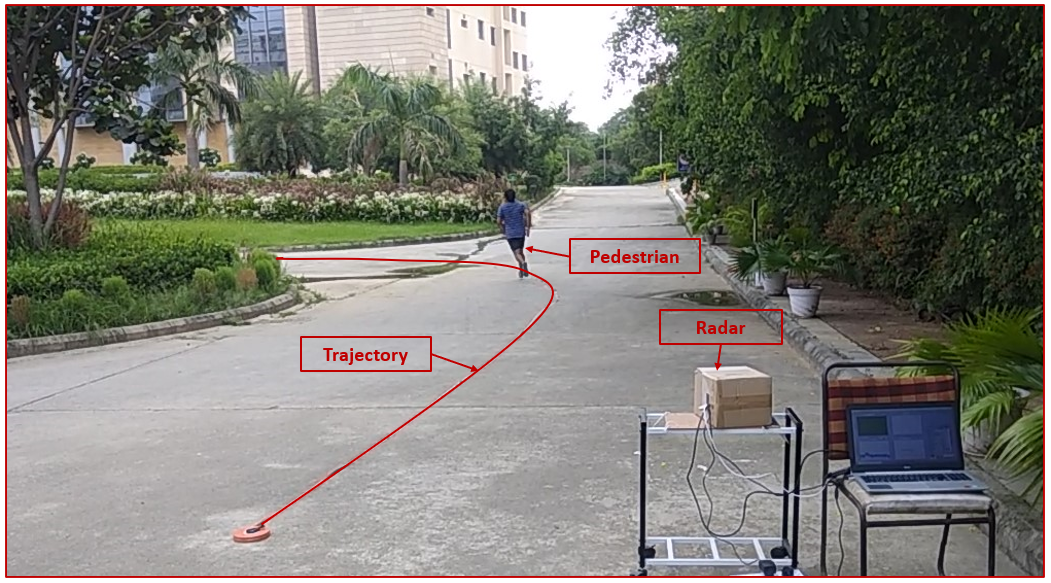}}
    \caption{Experimental setup for gathering ISAR images of (a) Radar hardware, (b) a small sized car, (c) pedestrian, undertaking a left turn along their trajectory.}
    \vspace{-2mm}
    \label{fig:Expermental_setup_left_turn_pedestrian}
\end{figure*}

\begin{figure}[htbp]
    \centering
    \subfigure[]{
    \includegraphics[width=1.5in, height = 1.5in]{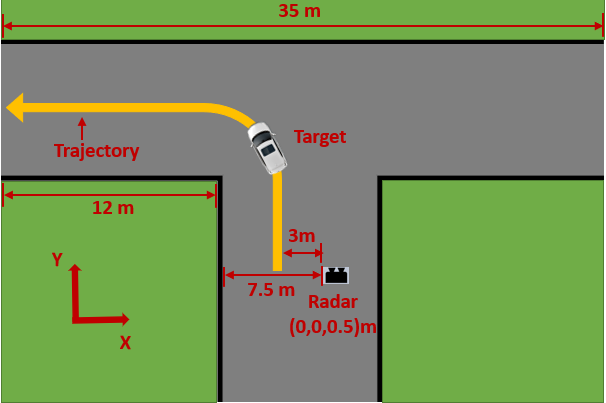}}
    \subfigure[]{
    \includegraphics[width = 1.5in, height = 1.5in]{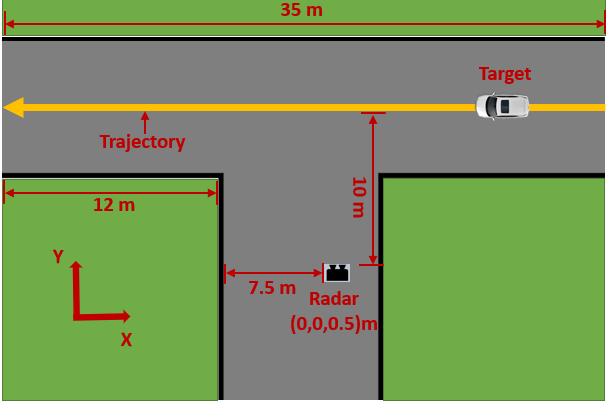}}
    \caption{Experimental setup for gathering ISAR images of (a) a left turn trajectory  before the radar, (b) tangential trajectory before the radar.}
    \vspace{-2mm}
    \label{fig:Expermental_setup_trajectories}
\end{figure}

In this section, we use Texas Instruments AWR-1843, a 77GHz millimeter-wave radar shown in Fig.\ref{fig:Expermental_setup_left_turn_pedestrian}a, to collect the experimental data, to validate the simulation results. We configured the radar to operate in an ultra-short range radar (USRR) mode. The configuration parameters that we used for our measurement are listed in Table.\ref{tab:RadarParameters}. We have chosen parameters to closely align with those used for the simulations in the previous section. The transmitted power from the radar is 14dBm and the noise floor of the receiver is approximately -110dBm. 
We considered two types of automotive targets as shown in Fig.\ref{fig:Expermental_setup_left_turn_pedestrian}b and c. The first is a small-sized car - Maruti Suzuki Celerio - of $3.695m \times 1.600m \times 1.560m$ dimensions. The second is a 1.8m tall pedestrian.
Each target moves independently in the channel. We considered two trajectories as shown in Fig.\ref{fig:Expermental_setup_trajectories}a and Fig.\ref{fig:Expermental_setup_trajectories}b. 
In the first trajectory, the car moves along a straight path a little left to the radar and then executes a left turn at an average speed of 10 kmph. In the second trajectory, the car moves along a tangential path before the radar from right to left at an average speed of 18 kmph. The same trajectories are repeated by a pedestrian with an average speed of 15 kmph in both cases.

\begin{figure*}[htbp]
    \centering
    \subfigure[]{
    \includegraphics[width=3.5in,height=1.75in]{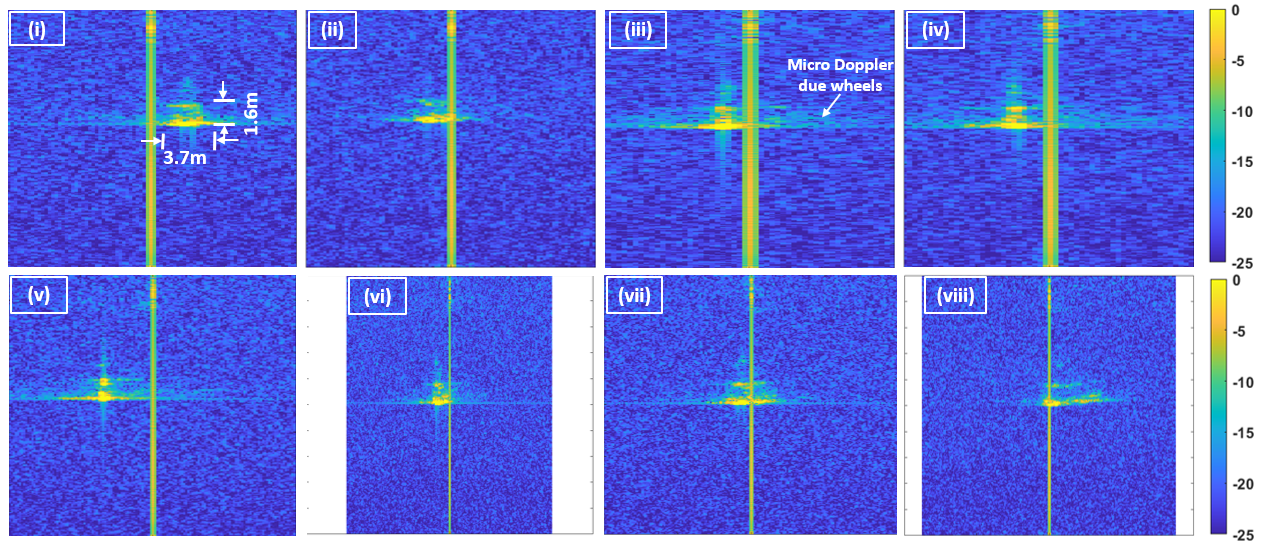}}
    \subfigure[]{
    \includegraphics[width=3.5in,height=1.75in]{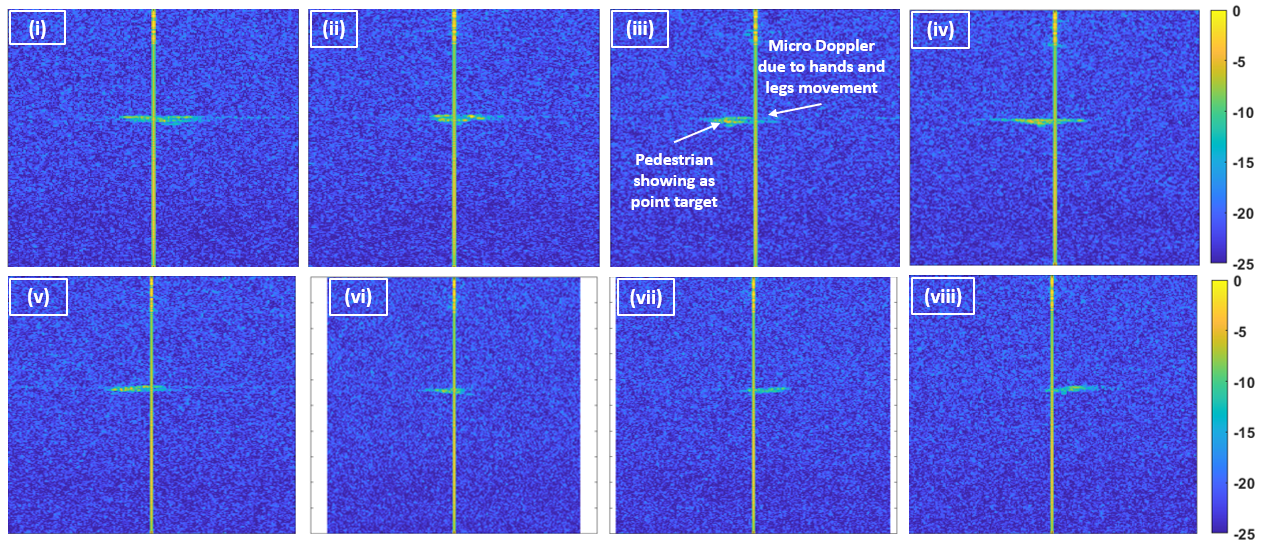}}
    \caption{ISAR images of (a) \emph{car}  (i-iv) along the tangential trajectory, (v-viii) along the left turn trajectory; (b) \emph{Pedestrian} (i-iv) along the tangential path with approx speed of 15 kmph  (v-viii) along the left turn path with approx speed of 15 kmph. The range span is 20m while the cross-range span varies from 10m to 20m. The dynamic range is 25dB.}
    \vspace{-2mm}
    \label{fig:ISAR images from capture data}
\end{figure*}

For both cases, we perform range compensation using estimates from the prior knowledge of the trajectory. Based on the motion of the target, we estimate the angular velocity of the target for every CPI. In real world scenarios, where the trajectory of the target and its velocity are not known a priori, a second radar or auxiliary sensor can be used to estimate the range and velocity as described in \cite{li2018wide}. Then, we perform matched filtering along the fast time and Doppler processing along the slow time to obtain ISAR images of the target for every CPI. The ISAR images of the car moving along the straight trajectory from left to right are shown in Fig.\ref{fig:ISAR images from capture data}a.i-iv. All the images have been normalized, and the dynamic range is 25dB. The images show the top-view dimensions of the vehicle along the range and cross-range dimensions. We observe the different positions of the car along the cross-range across the four images. From the dimensions in the figure, we observe that the car is oriented with its longer dimension along the cross-range. The images corresponding to the car making a left turn are shown in Fig.\ref{fig:ISAR images from capture data}a.v-viii. Here, we observe the longer dimension along the range in the first two figures and then the car orients along the cross-range in the last figure. In some frames, we observe a large spread along the cross-range dimension due to the micro-Dopplers arising due to the wheel motions. 

Fig.\ref{fig:ISAR images from capture data}b. shows the ISAR images of the pedestrian along the two trajectories. The top row shows the images when the pedestrian is moving along the first trajectory, while the bottom row shows the images corresponding to the second trajectory. The pedestrian, unlike the car, is very narrow along the range and cross-range. Hence, he appears like a point scatterer in the images. It is far more difficult to infer the trajectory of a pedestrian than a car from the ISAR images. The spread along the cross-range is due to the micro-Dopplers arising from the swinging motions of the arms and legs. 
The ISAR images generated from the measurement data show significant differences between the pedestrian and the car. Hence, they corroborate our earlier hypothesis that ISAR images offer distinctive features of targets that may be useful for classification purposes. 
\section{Classification Results}
\label{sec:Results}
In this section, we use classical machine learning techniques - support vector machine (SVM) and random forests (RF) - and more recent deep learning algorithms based on transfer learning - Alexnet and Googlenet - for classifying the five automotive targets on the basis of their ISAR images. We will examine the impact of noise and clutter and the volume of test and training data on the classification performance . 
\subsection{Effect of noise and clutter on classification performance}
\begin{figure}[htbp]
    \centering
\subfigure[]{
    \includegraphics[width=1.6in, height = 1.5in]{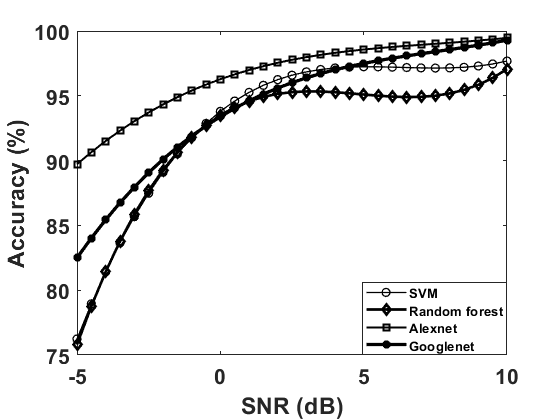}}
\subfigure[]{
    \includegraphics[width=1.6in,height=1.5in]{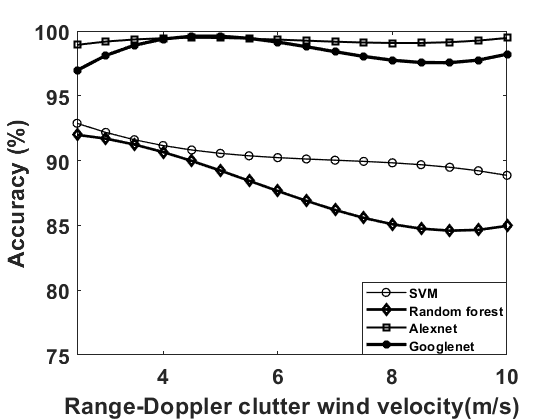}}
    \caption{Classification accuracy of automotive target ISAR images using SVM, random Forest and transfer learning algorithms based on Alexnet and Googlenet for (a) differing SNR values and (b) different range-Doppler clutter values. 70\% data used for training and remaining data used for validation and testing.}
    \vspace{-2mm}
    \label{fig:SNRvsAccuracy}
\end{figure}

Based on Table.\ref{tab:ISAR_database}, of the total volume of 14976 images for different SNR values, 70\% are used for training and 30\% for testing in the case of SVM and RF. In the case of Alexnet and Googlenet, we split the 30\% data that are not used for training between validation and testing. The resulting classification accuracy for different SNR values is shown in Fig.\ref{fig:SNRvsAccuracy}a.
We first observe that the classification accuracy for all algorithms is above 75\% even for low SNR of $-5dB$. The accuracy for SVM and RF are significantly poorer than those obtained from Alexnet and Googlenet at low SNR (-5dB). The classification accuracy improves for all cases as the SNR increases. 
The performances of Alexnet and Googlenet hold steady (above 80\% for Googlenet and 90\% for Alexnet) for all cases. 

We perform a similar study where we examine the effect of clutter on the classification performance of the ISAR images. The clutter along the range is modeled using a mean surface clutter coefficient. As the range increases, the area of coverage increases resulting in greater clutter. Wind gives rise to Doppler based clutter along the cross-range dimension. Higher wind velocities ($U$) give rise to greater clutter. Again, we have assumed a 70-30 split between training and test data for SVM and RF and a 70-15-15 split between training, validation, and test for Alexnet and Googlenet. We show the variation of the classification accuracy with respect to mean wind velocity in Fig.\ref{fig:SNRvsAccuracy}b. We observe that the classification performance for all the algorithms is fairly high (above 85\%) even for high values of clutter arising from high wind speeds (10m/s). The performance of the two transfer learning based algorithms (Alexnet and Googlenet) remains consistent even for the high values of clutter. On the other hand, we observe a slight deterioration in the performance of the two classical machine learning techniques with higher clutter values. The results indicate that the ISAR images offer highly discriminatory features for classification, even in the presence of high noise and clutter.
\subsection{Effect of test and training percentages}
In Table.\ref{tab:TrainingTestingMLResults}, we report the classification results for the four algorithms - SVM, RF, Alexnet and Googlenet - for different percentages of training, testing and validation data. 
\begin{table*}[htbp]
    \centering
\begin{tabular}{ |c|c|c|c|c|c|c| } 
\hline
Classifier & Training ($\%$) & Testing ($\%$) & Validation ($\%$)  & SNR & Range-Doppler Clutter & Combined \\
\hline
\multirow{3}{4em}{SVM} & 70 & 30 & - & 92.4 & 93.8 & 88.6\\
& 60 & 40 & - & 92.2 & 93.7 & 88.1 \\
& 50 & 50 & - & 92.3 & 93.3 & 87.7 \\
\hline
\multirow{3}{4em}{Random Forest} & 70 & 30 & - & 90.9 & 93.3 & 91.9\\
& 60 & 40 & - & 90.6 & 93.2 & 91.9 \\
& 50 & 50 & - & 90.9 & 92 & 90.8 \\
\hline
\multirow{3}{4em}{Alexnet} & 70 & 15 & 15 & 96.7 & 99.9 & 98.1\\
& 60 & 20 & 20 & 96.4 & 99.4 & 98.1 \\
& 50 & 25 & 25 & 95.5 & 97.7 & 97.3 \\
\hline
\multirow{3}{4em}{Googlenet} & 70 & 15 & 15 & 95.9 & 99.2 & 97.6\\
& 60 & 20 & 20 & 94.7 & 99.2 & 97.3 \\
& 50 & 25 & 25 & 94.4 & 98.6 & 97.2 \\
\hline
\end{tabular}
\caption{Classification of ISAR images using classical machine learning algorithms - SVM and random forest, and deep learning based algorithms - Alexnet and Googlenet}
\vspace{-2mm}
\label{tab:TrainingTestingMLResults}
\end{table*}
For each case, we have performed 5-fold cross-validation. We first consider the data that are just corrupted by noise (SNR varying from $+10dB$ to $-5dB$). Then we repeat the tests on data that are just corrupted by only clutter (wind velocities varying from 2.5 to 10m/s). Finally, we repeat the tests on data that are corrupted by both noise and clutter. We observe that the performances for all the cases are above 87\%. Since, the training data is very large; the algorithms perform well even when the training and testing are split evenly. We do not see a significant improvement in the performance with an increase in training data. The transfer learning algorithms like Alexnet and Googlenet perform very well even for low SNR and high clutter. 

In the following section, we present the confusion matrices obtained from the classification of data combining both noise and clutter. These results are obtained using 70\% training data. For each case, the rows show the true labels of the data, while columns show the labels of the predictions. We present three metrics with each confusion matrix. They are \emph{precision}, \emph{recall} and the $F_1$ score. Precision is defined as the ratio of the true positives (the highlighted number along the diagonal) over the sum of the true positives and false positives (column sum); while recall is defined as the ratio of the true positives over the sum of the true positives and false negatives (row sum). For each case, we also provide the $F_1$ score, which is defined as the harmonic mean of average precision and average recall,
\par\noindent\small
\begin{align}
\label{eq:F1 Score}
    F_1 = 2 \frac{{avg. precision}\times {avg. recall}}{{avg. precision}+{avg. recall}}.
\end{align}

The first confusion matrix is presented for the SVM in Table.\ref{tab:ConfusionMatrix}.
\begin{table*}[htbp]
    \centering
    \begin{tabular}{|c|c|c|c|c|c|c|c|}
         \hline
         \cellcolor{blue}SVM &  & \multicolumn{5}{c}{Predicted Labels} &\\
         \hline
         & Vehicle & Auto-rickshaw & Bicycle & Full size car & Mid size car & Truck & Recall  \\
         \hline
         \multirow{5}{4em}{True Labels}
         & Auto-rickshaw & \cellcolor{blue}1655 & 3 & 41 & 96 & 2 & 92.1 \\ 
         & Bicycle & 24 & \cellcolor{blue}1687 & 17 & 72 & 3 & 93.6 \\
         & Full size car & 48 & 39 & \cellcolor{blue}1478 & 193 & 34 & 82.5 \\
         & Mid size car & 111 & 112 & 146 & \cellcolor{blue}1463 & 2 & 79.8 \\
         & Truck & 14 & 2 & 56 & 15 & \cellcolor{blue}1673 & 95.1 \\
         \hline
         & Precision & 89.4 & 91.5 & 85.0 & 79.6 & 97.6&\\
         \hline
         \hline
         \cellcolor{blue}RF &  & \multicolumn{5}{c}{Predicted Labels} &\\
         \hline
         & Vehicle & Auto-rickshaw & Bicycle & Full size car & Mid size car & Truck & Recall  \\
         \hline
         \multirow{6}{4em}{True Labels}
         & Auto-rickshaw & \cellcolor{blue}1674 & 14 & 13 & 96 & 0 & 93.2 \\ 
         & Bicycle & 4 & \cellcolor{blue}1723 & 11 & 63 & 2 & 95.6 \\
         & Full size car & 29 & 34 & \cellcolor{blue}1577 & 145 & 7 & 88.0 \\
         & Mid size car & 65 & 130 & 67 & \cellcolor{blue}1570 & 2 & 85.6 \\
         & Truck & 3 & 4 & 24 & 18 & \cellcolor{blue}1711 & 97.2 \\
         \hline
         & Precision & 94.3 & 90.4 & 93.2 & 83.0 & 99.4 &\\
         \hline
         \hline
         \cellcolor{blue}Alexnet &  & \multicolumn{5}{c}{Predicted Labels} &\\
         \hline
         & Vehicle & Auto-rickshaw & Bicycle & Full size car & Mid size car & Truck & Recall  \\
         \hline
         \multirow{6}{4em}{True Labels}
         & Auto-rickshaw & \cellcolor{blue}895 & 1 & 0 & 2 & 0 & 99.7 \\ 
         & Bicycle & 0 & \cellcolor{blue}897 & 1 & 1 & 0 & 99.8 \\
         & Full size car & 3 & 1 & \cellcolor{blue}876 & 14 & 5 & 97.4 \\
         & Mid size car & 5 & 40 & 5 & \cellcolor{blue}849 & 0 & 94.4 \\
         & Truck & 2 & 0 & 4 & 0 & \cellcolor{blue}893 & 99.3 \\
         \hline
         & Precision & 98.9 & 95.5 & 98.9 & 98.0 & 99.4 &\\
         \hline
         \hline
         \cellcolor{blue}Googlenet &  & \multicolumn{5}{c}{Predicted Labels} &\\
         \hline
         & Vehicle & Auto-rickshaw & Bicycle & Full size car & Mid size car & Truck & Recall \\
         \hline
         \multirow{6}{4em}{True Labels}
         & Auto-rickshaw & \cellcolor{blue}881 & 0 & 8 & 7 & 2 & 98.1 \\ 
         & Bicycle & 0 & \cellcolor{blue}890 & 0 & 9 & 0 & 99.0 \\
         & Full size car & 0 & 3 & \cellcolor{blue}863 & 32 & 1 & 96.0 \\
         & Mid size car & 0 & 37 & 2 & \cellcolor{blue}860 & 0 & 95.7 \\
         & Truck & 0 & 0 & 8 & 1 & \cellcolor{blue}891 & 99.0\\
         \hline
        & Precision & 100 & 95.7 & 98.0 & 94.6 & 99.7 &\\
         \hline
    \end{tabular}
    \caption{Confusion matrices from SVM, RF, Alexnet and GoogleNet classifiers based on 70\% training, 15\% validation and 15\% test data.}
    \vspace{-2mm}
    \label{tab:ConfusionMatrix}
\end{table*}
We observe the least confusion for the truck. Due to its large dimensions and strong returns, the truck is rarely mistaken for any of the other targets or vice versa. Similarly, the bicycle is very small and hence not easily confused with the other targets. However, due to its weak returns and small size, sometimes, the bicycle is not easily discernible in noisy images. The two cars are often confused by each other due to the similarity in their dimensions, the number of wheels, and their strength of returns. The mid-size car, especially, has the lowest precision and recall because it is most similar to both full-size car and the auto-rickshaw. The $F_1$ score for SVM is 88.6\%. 

We observe a similar result for the RF classifier in the second matrix in Table.\ref{tab:ConfusionMatrix}. 
Again, the best precision and recall are observed for the large truck and the small bicycle. The results of the bicycle are slightly worse than the truck because of its weak returns, which get affected when the noise floor is high. Again, the two cars are confused by each other. However, this time, the results for the full-size car have significantly improved while those of the mid-size car have only slightly improved. There is a noticeable improvement in the performance of the Alexnet classifier, compared to the traditional machine learning algorithms for all cases, as reported the third matrix in Table.\ref{tab:ConfusionMatrix}.
Here, both the precision and recall for all the cases is above 95\%. Thus the two cars are no longer significantly confused by each other. The confusion between the auto-rickshaw and mid-size car has also substantially decreased. The same improvement is also observed for the Googlenet classifier as seen in the fourth matrix in Table.\ref{tab:ConfusionMatrix}.
Again, the accuracy are above 95\% for all five cases, both precision and recall. The $F_1$ scores for RF, Alexnet, and Googlenet are 92\%, 98.1\% and 97.6\%, respectively. 

\section{Conclusion}
\label{sec:Conclusion}
We have demonstrated an automotive radar simulation framework that incorporates radar scattering phenomenology of commonly found round vehicles as well as range-based surface clutter and Doppler based wind clutter and additive receiver noise. 
\balance
Using this simulation framework, we have demonstrated that high-resolution ISAR radar images, characterized by fine range and cross-range resolution, of dynamic automotive targets, can be generated with millimeter-wave automotive radars. A large database of over 30000 images has been publicly released to the radar community. The simulation framework has been verified through experimental data gathered with a real automotive millimeter-wave radar from Texas Instruments. These images provide meaningful information about the dimensions of the vehicle along the top-view as well as the number of wheels and the trajectory undertaken by the vehicle in the case of larger vehicles such as auto-rickshaws, cars, and trucks. Smaller targets such as pedestrians and bicycles, on the other hand, more closely resemble single point scatterers. These images facilitate highly accurate target classification (above 90\%) with both traditional machine learning techniques as well as the more recent deep neural networks. 

\bibliographystyle{IEEEtran}
\bibliography{main}

\begin{thebibliography}{10}
\providecommand{\url}[1]{#1}
\csname url@samestyle\endcsname
\providecommand{\newblock}{\relax}
\providecommand{\bibinfo}[2]{#2}
\providecommand{\BIBentrySTDinterwordspacing}{\spaceskip=0pt\relax}
\providecommand{\BIBentryALTinterwordstretchfactor}{4}
\providecommand{\BIBentryALTinterwordspacing}{\spaceskip=\fontdimen2\font plus
\BIBentryALTinterwordstretchfactor\fontdimen3\font minus
  \fontdimen4\font\relax}
\providecommand{\BIBforeignlanguage}[2]{{%
\expandafter\ifx\csname l@#1\endcsname\relax
\typeout{** WARNING: IEEEtran.bst: No hyphenation pattern has been}%
\typeout{** loaded for the language `#1'. Using the pattern for}%
\typeout{** the default language instead.}%
\else
\language=\csname l@#1\endcsname
\fi
#2}}
\providecommand{\BIBdecl}{\relax}
\BIBdecl

\bibitem{rasshofer2005automotive}
R.~H. Rasshofer and K.~Gresser, ``Automotive radar and lidar systems for next
  generation driver assistance functions.'' \emph{Advances in Radio Science},
  vol.~3, 2005.

\bibitem{schneider2005automotive}
M.~Schneider, ``Automotive radar-status and trends,'' in \emph{German microwave
  conference}, 2005, pp. 144--147.

\bibitem{hasch2012millimeter}
J.~Hasch, E.~Topak, R.~Schnabel, T.~Zwick, R.~Weigel, and C.~Waldschmidt,
  ``Millimeter-wave technology for automotive radar sensors in the 77 ghz
  frequency band,'' \emph{IEEE Transactions on Microwave Theory and
  Techniques}, vol.~60, no.~3, pp. 845--860, 2012.

\bibitem{fleming2012recent}
B.~Fleming, ``Recent advancement in automotive radar systems [automotive
  electronics],'' \emph{IEEE Vehicular Technology Magazine}, vol.~7, no.~1, pp.
  4--9, 2012.

\bibitem{jouny1993classification}
I.~Jouny, F.~Garber, and S.~Ahalt, ``Classification of radar targets using
  synthetic neural networks,'' \emph{IEEE Transactions on Aerospace and
  Electronic Systems}, vol.~29, no.~2, pp. 336--344, 1993.

\bibitem{chiang2000model}
H.-C. Chiang, R.~L. Moses, and L.~C. Potter, ``Model-based classification of
  radar images,'' \emph{IEEE Transactions on Information Theory}, vol.~46,
  no.~5, pp. 1842--1854, 2000.

\bibitem{bilik2006gmm}
I.~Bilik, J.~Tabrikian, and A.~Cohen, ``Gmm-based target classification for
  ground surveillance doppler radar,'' \emph{IEEE Transactions on Aerospace and
  Electronic Systems}, vol.~42, no.~1, pp. 267--278, 2006.

\bibitem{kim2009human}
Y.~Kim and H.~Ling, ``Human activity classification based on micro-doppler
  signatures using a support vector machine,'' \emph{IEEE Transactions on
  Geoscience and Remote Sensing}, vol.~47, no.~5, pp. 1328--1337, 2009.

\bibitem{kim2015human}
Y.~Kim and T.~Moon, ``Human detection and activity classification based on
  micro-doppler signatures using deep convolutional neural networks,''
  \emph{IEEE Geoscience and Remote Sensing Letters}, vol.~13, no.~1, pp. 8--12,
  2015.

\bibitem{vishwakarma2018dictionary}
S.~Vishwakarma and S.~S. Ram, ``Dictionary learning with low computational
  complexity for classification of human micro-dopplers across multiple carrier
  frequencies,'' \emph{IEEE Access}, vol.~6, pp. 29\,793--29\,805, 2018.

\bibitem{seyfiouglu2018deep}
M.~S. Seyfio{\u{g}}lu, A.~M. {\"O}zbayo{\u{g}}lu, and S.~Z. G{\"u}rb{\"u}z,
  ``Deep convolutional autoencoder for radar-based classification of similar
  aided and unaided human activities,'' \emph{IEEE Transactions on Aerospace
  and Electronic Systems}, vol.~54, no.~4, pp. 1709--1723, 2018.

\bibitem{fioranelli2015classification}
F.~Fioranelli, M.~Ritchie, and H.~Griffiths, ``Classification of unarmed/armed
  personnel using the netrad multistatic radar for micro-doppler and singular
  value decomposition features,'' \emph{IEEE Geoscience and Remote Sensing
  Letters}, vol.~12, no.~9, pp. 1933--1937, 2015.

\bibitem{prophet2018pedestrian}
R.~Prophet, M.~Hoffmann, A.~Ossowska, W.~Malik, C.~Sturm, and M.~Vossiek,
  ``Pedestrian classification for 79 ghz automotive radar systems,'' in
  \emph{2018 IEEE Intelligent Vehicles Symposium (IV)}.\hskip 1em plus 0.5em
  minus 0.4em\relax IEEE, 2018, pp. 1265--1270.

\bibitem{khomchuk2016pedestrian}
P.~Khomchuk, I.~Stainvas, and I.~Bilik, ``Pedestrian motion direction
  estimation using simulated automotive mimo radar,'' \emph{IEEE Transactions
  on Aerospace and Electronic Systems}, vol.~52, no.~3, pp. 1132--1145, 2016.

\bibitem{ritchie2016multistatic}
M.~Ritchie, F.~Fioranelli, H.~Borrion, and H.~Griffiths, ``Multistatic
  micro-doppler radar feature extraction for classification of unloaded/loaded
  micro-drones,'' \emph{IET Radar, Sonar \& Navigation}, vol.~11, no.~1, pp.
  116--124, 2016.

\bibitem{molchanov2013classification}
P.~Molchanov, K.~Egiazarian, J.~Astola, R.~Harmanny, and J.~De~Wit,
  ``Classification of small uavs and birds by micro-doppler signatures,'' in
  \emph{2013 European Radar Conference}.\hskip 1em plus 0.5em minus 0.4em\relax
  IEEE, 2013, pp. 172--175.

\bibitem{bilik2007radar}
I.~Bilik and J.~Tabrikian, ``Radar target classification using doppler
  signatures of human locomotion models,'' \emph{IEEE Transactions on Aerospace
  and Electronic Systems}, vol.~43, no.~4, pp. 1510--1522, 2007.

\bibitem{chen2016target}
S.~Chen, H.~Wang, F.~Xu, and Y.-Q. Jin, ``Target classification using the deep
  convolutional networks for sar images,'' \emph{IEEE Transactions on
  Geoscience and Remote Sensing}, vol.~54, no.~8, pp. 4806--4817, 2016.

\bibitem{wagner2016sar}
S.~A. Wagner, ``Sar atr by a combination of convolutional neural network and
  support vector machines,'' \emph{IEEE Transactions on Aerospace and
  Electronic Systems}, vol.~52, no.~6, pp. 2861--2872, 2016.

\bibitem{lin2017deep}
Z.~Lin, K.~Ji, M.~Kang, X.~Leng, and H.~Zou, ``Deep convolutional highway unit
  network for sar target classification with limited labeled training data,''
  \emph{IEEE Geoscience and Remote Sensing Letters}, vol.~14, no.~7, pp.
  1091--1095, 2017.

\bibitem{ram2015high}
S.~S. Ram and A.~Majumdar, ``High-resolution radar imaging of moving humans
  using doppler processing and compressed sensing,'' \emph{IEEE Transactions on
  Aerospace and Electronic Systems}, vol.~51, no.~2, pp. 1279--1287, 2015.

\bibitem{chen2014inverse}
V.~C. Chen, \emph{Inverse Synthetic Aperture Radar Imaging; Principles}.\hskip
  1em plus 0.5em minus 0.4em\relax Institution of Engineering and Technology,
  2014.

\bibitem{martorella2009automatic}
M.~Martorella, E.~Giusti, A.~Capria, F.~Berizzi, and B.~Bates, ``Automatic
  target recognition by means of polarimetric isar images and neural
  networks,'' \emph{IEEE Transactions on Geoscience and Remote Sensing},
  vol.~47, no.~11, pp. 3786--3794, 2009.

\bibitem{vespe2007automatic}
M.~Vespe, C.~Baker, and H.~Griffiths, ``Automatic target recognition using
  multi-diversity radar,'' \emph{IET Radar, Sonar \& Navigation}, vol.~1,
  no.~6, pp. 470--478, 2007.

\bibitem{park2011construction}
S.-H. Park, M.-G. Joo, and K.-T. Kim, ``Construction of isar training database
  for automatic target recognition,'' \emph{Journal of Electromagnetic Waves
  and Applications}, vol.~25, no. 11-12, pp. 1493--1503, 2011.

\bibitem{danylov2010terahertz}
A.~A. Danylov, T.~M. Goyette, J.~Waldman, M.~J. Coulombe, A.~J. Gatesman, R.~H.
  Giles, X.~Qian, N.~Chandrayan, S.~Vangala, K.~Termkoa \emph{et~al.},
  ``Terahertz inverse synthetic aperture radar (isar) imaging with a quantum
  cascade laser transmitter,'' \emph{Optics express}, vol.~18, no.~15, pp.
  16\,264--16\,272, 2010.

\bibitem{kulpa2013experimental}
J.~S. Kulpa, M.~Malanowski, D.~Gromek, P.~Samczy{\'n}sk, K.~Kulpa, and
  A.~Gromek, ``Experimental results of high-resolution isar imaging of
  ground-moving vehicles with a stationary fmcw radar,'' \emph{International
  Journal of Electronics and Telecommunications}, vol.~59, no.~3, pp. 293--299,
  2013.

\bibitem{li2018wide}
C.~Li and H.~Ling, ``Wide-angle, ultra-wideband isar imaging of vehicles and
  drones,'' \emph{Sensors}, vol.~18, no.~10, p. 3311, 2018.

\bibitem{essen2008high}
H.~Essen, M.~Hagelen, W.~Johannes, R.~Sommer, A.~Wahlen, M.~Schlechtweg, and
  A.~Tessmann, ``High resolution millimetre wave measurement radars for ground
  based sar and isar imaging,'' in \emph{2008 IEEE Radar Conference}.\hskip 1em
  plus 0.5em minus 0.4em\relax IEEE, 2008, pp. 1--5.

\bibitem{pandey2020database}
N.~Pandey, G.~Duggal, and S.~S. Ram, ``Database of simulated inverse synthetic
  aperture radar images for short range automotive radar,'' in \emph{2020 IEEE
  International Radar Conference (RADAR)}.\hskip 1em plus 0.5em minus
  0.4em\relax IEEE, 2020, pp. 238--243.

\bibitem{skolnik1980introduction}
M.~I. Skolnik \emph{et~al.}, \emph{Introduction to radar systems}.\hskip 1em
  plus 0.5em minus 0.4em\relax McGraw-hill New York, 1980, vol.~3.

\bibitem{kulemin2003millimeter}
G.~P. Kulemin, \emph{Millimeter-wave radar targets and clutter}.\hskip 1em plus
  0.5em minus 0.4em\relax Artech House, 2003.

\bibitem{suykens1999least}
J.~A. Suykens and J.~Vandewalle, ``Least squares support vector machine
  classifiers,'' \emph{Neural processing letters}, vol.~9, no.~3, pp. 293--300,
  1999.

\bibitem{liaw2002classification}
A.~Liaw, M.~Wiener \emph{et~al.}, ``Classification and regression by
  randomforest,'' \emph{R news}, vol.~2, no.~3, pp. 18--22, 2002.

\bibitem{torrey2010transfer}
L.~Torrey and J.~Shavlik, ``Transfer learning,'' in \emph{Handbook of research
  on machine learning applications and trends: algorithms, methods, and
  techniques}.\hskip 1em plus 0.5em minus 0.4em\relax IGI global, 2010, pp.
  242--264.

\bibitem{deep2020radar}
Y.~Deep, P.~Held, S.~S. Ram, D.~Steinhauser, A.~Gupta, F.~Gruson, A.~Koch, and
  A.~Roy, ``Radar cross-sections of pedestrians at automotive radar frequencies
  using ray tracing and point scatterer modelling,'' \emph{IET Radar, Sonar \&
  Navigation}, 2020.

\bibitem{ram2010simulation}
S.~S. Ram, C.~Christianson, Y.~Kim, and H.~Ling, ``Simulation and analysis of
  human micro-dopplers in through-wall environments,'' \emph{IEEE Transactions
  on Geoscience and remote sensing}, vol.~48, no.~4, pp. 2015--2023, 2010.

\bibitem{duggal2019micro}
G.~Duggal, K.~V. Mishra, and S.~S. Ram, ``Micro-doppler and micro-range
  detection via doppler-resilient 802.11ad-based vehicle-to-pedestrian radar,''
  in \emph{IEEE Radar Conference 2019}, 2019, pp. 1--6.

\bibitem{duggal2020doppler}
G.~Duggal, S.~Vishwakarma, K.~V. Mishra, and S.~S. Ram, ``Doppler-resilient
  802.11ad-based ultra-short range automotive joint radar-communications
  system,'' \emph{IEEE Transactions on Aerospace and Electronic Systems}, 2020.

\bibitem{ruck1970radar}
G.~T. Ruck, D.~E. Barrick, W.~D. Stuart, and C.~K. Krichbaum, \emph{Radar cross
  section handbook}.\hskip 1em plus 0.5em minus 0.4em\relax Plenum press New
  York, 1970, vol.~1.

\bibitem{richards2005fundamentals}
M.~A. Richards, \emph{Fundamentals of radar signal processing}.\hskip 1em plus
  0.5em minus 0.4em\relax Tata McGraw-Hill Education, 2005.

\bibitem{ulaby2019handbook}
F.~Ulaby, M.~C. Dobson, and J.~L. {\'A}lvarez-P{\'e}rez, \emph{Handbook of
  radar scattering statistics for terrain}.\hskip 1em plus 0.5em minus
  0.4em\relax Artech House, 2019.

\bibitem{king1970terrain}
H.~King, C.~Zamites, D.~Snow, and R.~Colliton, ``Terrain backscatter
  measurements at 40 to 90 ghz,'' \emph{IEEE Transactions on Antennas and
  Propagation}, vol.~18, no.~6, pp. 780--784, 1970.

\bibitem{pandey2012study}
P.~Pandey, D.~Kumar, A.~Prakash, J.~Masih, M.~Singh, S.~Kumar, V.~K. Jain, and
  K.~Kumar, ``A study of urban heat island and its association with particulate
  matter during winter months over delhi,'' \emph{Science of the Total
  Environment}, vol. 414, pp. 494--507, 2012.

\end{thebibliography}
\end{document}